\documentclass{jfm}
\usepackage{amsmath}
\usepackage{amssymb}
\usepackage{graphicx}
\usepackage{tikz}

\usepackage[caption=false,labelformat=simple]{subfig}
\usepackage{floatrow}
\newcommand{\drawline}[2]{\raisebox{2.5pt}{\vbox{\hrule width #1 pt height #2 pt}}}
\newcommand{\spacce}[1]{\hspace{#1pt}}
\newcommand{\solid}{\drawline{24}{0.5}\spacce{2}}

\newcommand{\pard}[2]{\frac{\partial #2}{\partial #1}}
\newcommand{\pardt}[1]{\frac{\partial #1}{\partial t}}

\newcommand{\pardz}[1]{\frac{\partial #1}{\partial z}}
\newcommand{\pardr}[1]{\frac{\partial #1}{\partial r}}
\newcommand{\pardpsi}[1]{\frac{\partial #1}{\partial \psi}}

\newcommand{\pardzz}[1]{\frac{\partial^2 #1}{\partial z^2}}
\newcommand{\pardrr}[1]{\frac{\partial^2 #1}{\partial r^2}}
\newcommand{\pardpsipsi}[1]{\frac{\partial^2 #1}{\partial \psi^2}}

\newcommand{\eavg}{\langle \mathbf{e} \rangle}
\newcommand{\esth}[1]{\langle e_{#1} \rangle}
\newcommand{\eo}[1]{{\langle e_{#1} \rangle}_0}
\newcommand{\diffusion}{\mathsfbi{D}}
\newcommand\ddfrac[2]{\frac{\displaystyle #1}{\displaystyle #2}}
\newcommand\Ri{\mbox{\textit{Ri}}}  

\usepackage[final]{changes} 

\begin{document}

\newtheorem{lemma}{Lemma}
\newtheorem{corollary}{Corollary}

\shorttitle{Downflowing gyrotactic suspension} 

\shortauthor{L. Fung, R. N. Bearon and Y. Hwang} 

\title{Bifurcation and stability of downflowing gyrotactic micro-organism suspensions  \\ in a vertical pipe}

\author
 {
 Lloyd Fung\aff{1}
  \corresp{\email{lloyd.fung@imperial.ac.uk}},
  Rachel N. Bearon\aff{2}
  \and 
  Yongyun Hwang\aff{1}
  }

\affiliation
{
\aff{1}
Department of Aeronautics, Imperial College London, London, SW7 2AZ, UK
\aff{2}
Department of Mathematical Sciences, University of Liverpool, Liverpool L69 7ZL, UK
}

\maketitle

\begin{abstract}
In the experiment that first demonstrated gyrotactic behaviour of bottom-heavy swimming microalgae (e.g. \textit{Chlamydomonas}), Kessler (\textit{Nature}, vol. 313, 1985, pp. 218-220)  showed that a beam-like structure, often referred to as a gyrotactic plume, would spontaneously appear from a suspension of gyrotactic swimmers in a downflowing pipe. Such a plume is prone to an instability to form blips.
This work models the gyrotactic plume as a steady parallel basic state and its subsequent breakdown into blips as an instability, employing both the Generalised Taylor Dispersion (GTD) theory and the Fokker-Planck model for comparison.
Upon solving for the basic state, it is discovered that the steady plume solution undergoes sophisticated bifurcations.
When there is no net flow, there exists a non-trivial solution of the plume structure other than the stationary uniform suspension, stemming from a transcritical bifurcation with the average cell concentration.
When a net downflow is prescribed, there exists a cusp bifurcation. Furthermore, there is a critical concentration, at which the cell concentration at the centre would blow up for the GTD model. 
The subsequent stability analysis using the steady plume solution shows that the Fokker-Planck model is inconsistent with what was experimentally observed, as it predicts stabilisation of axisymmetric blips at high concentration of the plume and destabilisation of the first non-axisymmetric mode at low flow rates. 
\end{abstract}

\section{Introduction\label{sec:Introduction}}
In the natural environment and industrial applications, the motility (or swimming motion) of micro-organisms plays an important role in their transport processes. These motile micro-organisms often swim towards light, food, oxygen or against gravity, as it would offer a better living condition. These stimulus-guided motility of micro-organisms are called `taxis'. In a suspension of these micro-organisms, such taxes can significantly impact on how they distribute themselves both spatially and temporally as well as on rheological properties. \added{For example, in a shallow suspension of swimming algae where bioconvective patterns are often observed, light, oxygen and gravity are known to suppress or encourage the pattern formation \citep{Bees2020}.}

Gyrotaxis is an example of such taxes, and is seen in some unicellular species of algae, such as \emph{Chlamydomonas}, \emph{Dunaliella} and \emph{Heterosigma}, which are bottom-heavy (i.e. the centre of gravity is offset from the centre of buoyancy). These micro-organisms typically experience a gravitational torque due to the bottom-heaviness, and it makes them orient upwards in the absence of flow. However, in the presence of a vortical flow, their orientation deviates from the vertical due to the viscous torque from the flow vorticity \added{\citep{Kessler1985a}}. For example, in a vertical pipe with a downward flow, the balance between the viscous torque and the gravitational torque creates a swimming-oriented flux of cells towards the centre, resulting in their accumulation along the centerline of the pipe. The column of such accumulated cells, often called a gyrotactic plume, can further accelerate the flow at the centreline due to the negative buoyancy force exerted by the cells, thereby further increasing the shear and attracting even more cells towards itself \added{\citep{Kessler1986}}. 

The gyrotactic plume was first documented in a series of experimental studies carried out with vertical pipe flows by \cite{Kessler1984, Kessler1985b, Kessler1985a, Kessler1986}, who coined the term `gyrotaxis' to describe how the orientation of a bottom-heavy cell is influenced by the vorticity in the surrounding flow and the gravity. Further to the formation of a gyrotatic plume along the pipe centreline in the downward flow, he observed that, under certain conditions, the plume could experience an instability, which subsequently breaks it down into multiple blips. The blips are the localised regions of a higher density of the cells, and are spontaneously formed along the gyrotactic plume at the late stage of the instability. In general, they are more pronounced when the background concentration is high. They are also almost uniformly spaced vertically and remain axisymmetric \citep{Kessler1986,Denissenko2007}. 

\cite{Kessler1986} originally modelled the spatial distribution of the cells with a simple advection-diffusion-based transport equation for cell concentration to understand his experimental observations\added{, where the horizontal advection velocity was modelled to be proportional to ambient vorticity and the diffusivity was assumed to be constant and isotropic}. Although the model is somewhat primitive, it enabled him to subsequently find an analytic solution of the plume structure for the special case where the imposed pressure gradient is zero. Since the pioneering work of Kessler, numerous efforts have been made to improve the description for the transport of cell concentration. \cite{Pedley1990} introduced the first model accounting for the random walk of individual cells, in which the swimming orientation of an individual cell was described by \added{the} \deleted{a stochastically-driven} Jeffery's equation \citep{Jeffery1922,Hinch1972a,Hinch1972} \added{with a superposed white noise for rotational random walk}. A probability density function (p.d.f) for the cell orientation was obtained by formulating a Fokker-Plank equation, and the diffusivity was approximated from the related correlation of the cell orientation vector. \cite{Hill2002} and \cite{Manela2003} later proposed the use of generalised Taylor dispersion (GTD) theory, which approximates the diffusivity from the p.d.f of a single tracer particle defined in both orientation and physical space. 
\cite{Bees2010a} and \cite{Bearon2012} recently employed the GTD model to study the dispersion of gyrotactic cells in a downward pipe flow. In particular, \cite{Bearon2012} showed that the prediction of the GTD model on the cell-concentration distribution is significantly different from that of the model of \cite{Pedley1990}. Importantly, a recent numerical study by \cite{Croze2013} further showed that the GTD model provides a much more accurate prediction for the cell distribution obtained from individual-based simulation than the one by \cite{Pedley1990} especially when the shear (or vorticity) rate of the surrounding flow is high. \added{Recently, \cite{Jiang2020} have also demonstrated the superiority of the GTD model without making any approximation to the Smoluchowski equation. These observations were} \deleted{This observation was further} supported by the recent experimental result in \cite{Croze2017}. 

In the light of emerging evidence supporting the use of the GTD model at least for unidirectional flows \cite[for this issue, see also][]{Bearon2011}, the present study aims to study the emergence of the blips observed in the original pipe-flow experiment of \cite{Kessler1986} using the GTD model. Recently, the stability of gyrotactic plumes emerging in a downward plane channel flow was studied by \cite{HP2014b} using a model basically identical to the one by \cite{Pedley1990}. Interestingly, they reported the emergence of a varicose-type instability mode sharing some similarities with the blips observed in pipe flow (e.g. spacing between blips). Despite the encouraging observation, no such analysis is available for pipe flows, and therefore it is not possible to make any direct comparison with the previous experiments \cite[]{Kessler1986,Denissenko2007,Croze2017}. 
Nevertheless, bifurcation and stability of any gyrotactic micro-organism suspensions have only been studied either using the early primitive model of \cite{Pedley1988} \cite[e.g.][]{Hill1989} or using the model of \cite{Pedley1990} \cite[e.g.][]{Bees1997,Pedley2010a,HP2014a,HP2014b,Maretvadakethope2019}. For this reason, the issue of how important having an accurate cell-transport model is for the prediction of the pattern-forming motions in the suspension remains completely elusive.

\begin{figure}
    \centering
    \includegraphics[width=0.45\columnwidth]{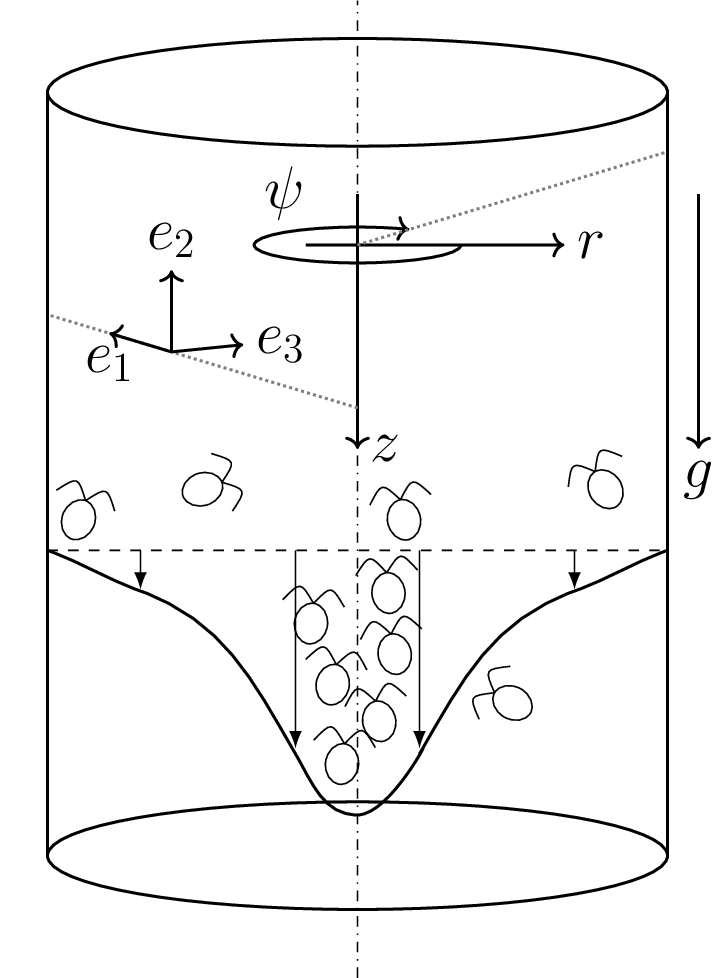}
    \caption{Schematic diagram of the flow configuration and the frames of reference. Here, the stability analysis is performed on the global frame of reference $\mathbf{x}=(r,\psi,z)$, but the cell swimming direction is calculated in the local frame of reference $\mathbf{e}=(e_1,e_2,e_3)$.}
    \label{fig:1}
\end{figure}

The objective of the present study is to extensively study the bifurcation and stability of downflowing suspensions in a vertical pipe using the GTD model. Unlike the channel flow studied by \cite{HP2014b}, in this set up we shall see the emergence of bistability and the related hysteresis in the bifurcation of the basic state, the possibility of which was previously conjectured by \cite{Bees2010a}. Further to this, a particular emphasis of the present study is given to an extensive and comparative assessment of the model by \cite{Pedley1990} and the GTD model for studying bifurcation and stability of the suspension in a downward pipe flow. From this, we shall see that the prediction of the GTD model offers a much more realistic description for the experimental observation on the blip instability than that of the model by \cite{Pedley1990}, highlighting the need for having a correct cell transport model to accurately describe pattern-forming fluid motions in suspensions of gyrotactic micro-organisms. 

This paper is organised as follows. In \S \ref{sec:Problem_Formulation}, the equations of motion are introduced and formulated for linear stability analysis with the model of \cite{Pedley1990} and the GTD model. In \S \ref{sec:Results_basic_state}, the bifurcation of the basic state given in the form of a steady axisymmetric and axially uniform plume is first shown with particular emphasis on their bistable nature in a range of the given parameters. The instability of these basic states is subsequently studied in \S \ref{sec:Results_stability}. The implication of these findings is discussed in \S \ref{sec:Discussion} from both experimental and theoretical perspectives. The paper is concluded with a brief discussion on the shortcoming of the GTD model with an outlook to improve this model.

\section{Problem formulation \label{sec:Problem_Formulation}}
\subsection{Equations of fluid motion}
We consider a downward fluid flow in a cylindrical vertical pipe, in which a puller type of spherical gyrotactic micro-organisms are suspended (e.g. \emph{Chlamydomonas nivalis}, \emph{Dunaliella} and \emph{Volvox}). We express the position of a point in space in terms of cylindrical co-ordinates $r^*$, $\psi^*$ and $z^*$, the radial, azimuthal and streamwise (or axial) co-ordinates, and the corresponding unit vectors $\mathbf{i}$, $\mathbf{j}$ and $\mathbf{k}$. The fluid has density $\rho^*$ and kinematic viscosity $\nu^*$, and the gravity is in the same direction as $z^*$. The local coordinates of $\mathbf{e}=(e_1,e_2,e_3)$ are introduced for the swimming orientation of the cells, and are defined on the surface of unit sphere (i.e. $\|\mathbf{e}\|=1$), where $e_1$, $e_2$ and $e_3$ indicate the radial, upward ($-z^*$) and azimuthal components, respectively. Both of the coordinate systems and the flow geometry are sketched in figure \ref{fig:1}. We will also repeat some of the key assumptions made in \cite{HP2014a,HP2014b}, where a more detailed discussion and justifications on the assumptions in the present study can be found. The fluid motion is described by the following equations:
\begin{equation}
    \bnabla^* \bcdot \mathbf{u}^*  = 0, \label{eq:dim_contin}
\end{equation}
\begin{equation}
    \pard{t^*}{\mathbf{u}^*} + (\mathbf{u}^* \bcdot \bnabla^*) \mathbf{u}^* = -\frac{1}{\rho^*} \bnabla^* p^* + \nu^* \nabla^{*2} \mathbf{u}^* + n^* \upsilon^* g'^* \mathbf{k}+\bnabla^* \cdot \boldsymbol{\Sigma}_p^*.  \label{eq:dim_vel}
\end{equation}
Here, the velocity, $\mathbf{u}^*=(v^*,w^*,u^*)$, is resolved into the cylindrical components, $p^*$ is the pressure, $n^*$ the cell number density, $\upsilon^*$ the volume of a single cell and $g'^*=g ^*\Delta \rho^* /  \rho^*$ the reduced gravity, where $g^*$ and $\Delta \rho^*$ are the gravitational acceleration and the density difference between the microorganism and the fluid, respectively.
Following \cite{HP2014b}, we neglect the stresslet term $\bnabla^* \cdot \boldsymbol{\Sigma}_p^*$, which would be dominantly contributed by the locomotion of the suspended cells \citep{Pedley1990}. For the `puller'-type swimmers, the stresslet term $\boldsymbol{\Sigma}_p^*$ is not, in general, responsible for the generation of any instabilities \citep{Saintillan2007,Saintillan2008,Pedley2010a}. 
This term is also negligible compared to the negative buoyancy \citep{Pedley1990}. 
We will also neglect any near-field cell-cell interactions \citep{Kessler1992,Ishikawa2007} based on the assumption that the suspension is dilute enough. The validity of the assumption of dilute suspension here will be revisited later in \S \ref{sec:Validity}. 

\subsection{Models for cell transport\label{subsec:GTD_FK}}

Now, we introduce the transport equation for the cell concentration given in the form of an advection-diffusion equation: 
\begin{equation}
    \pard{t^*}{n^*} + \bnabla^* \bcdot [(\mathbf{u}^* + V^*_c \eavg ) n^*]  =  \bnabla^* \bcdot (\diffusion_m^* \bcdot \bnabla^* n^*), \label{eq:dim_n_gen}
\end{equation}
where $\mathbf{u}^* + V^*_c \eavg$ is the advection velocity, in which $V^*_c$ is the cell swimming velocity assumed to be constant and $\eavg$ represents the locally ensemble-averaged swimming orientation vector, and ${\diffusion}_m^*$ is the `effective' diffusivity tensor, the detailed definition of which will be introduced below with the subscript $m$. Here, we note that the diffusivity tensor ${\diffusion}_m^*$ is designed to model the spatio-temporal dispersion of the cells phenomenologically.  It models the random walk originating from the cell motility and the random cell orientation. Whilst it can also incorporate translational Brownian motion, because the typical size of the cells of interest in the present study (e.g. \emph{Chlamydomonas}) is too large to experience such a thermal fluctuation, we neglect this here \citep[see][]{PK1992,Saintillan2018}.

Based on (\ref{eq:dim_n_gen}), we now introduce two models for $\eavg$ and ${\diffusion}_m^*$. The first model, often called the Fokker-Planck model, was first introduced by \cite{Pedley1990}. The second model is based on the GTD theory, first proposed by \cite{Brenner1980}. This theory was previously applied to gyrotactic micro-organisms by \cite{Hill2002} and \cite{Manela2003}. For simplicity, hereafter, we shall refer to the Fokker-Planck model as model F, and the model based on GTD theory as model G, respectively. The behaviours of $\eavg$ and ${\diffusion}_m^*$ with respect to external velocity gradient in the two models have been extensively discussed in several previous studies  \citep{Manela2003,Bearon2012,Croze2013,Croze2017}.
Therefore, in the following sections, we will introduce them only briefly.

\subsubsection{Model F}
In model F, each cell is modelled as a motile spheroid performing a random rotational motion with gyrotaxis. For this purpose, a probability density function (p.d.f.) $f$ of the cell swimming orientation is introduced to describe the random rotation in the $\mathbf{e}$-space, which  satisfies the following steady Fokker-Planck equation \citep{Pedley1990}:
\begin{subequations}
\begin{equation}
    \bnabla_e \bcdot \left[ \mathbf{\dot{e}}^* f \right] =   D_R^* \bnabla^2_e f, \label{eq:pdf_gen}
\end{equation}
subject to 
\begin{equation}
    \int_{\|\mathbf{e}\|=1} f(\mathbf{e}) d^2\mathbf{e}=1,\label{eq:f_normalisation}
\end{equation}
and $\mathbf{\dot{e}}^*$ is given by the Jeffrey's equation \citep{Jeffery1922,Hinch1972a, Hinch1972}
\begin{equation}
\mathbf{\dot{e}}^*  =  \frac{1}{2\beta^*}[-\mathbf{k}+(\mathbf{k} \bcdot \mathbf{e})\mathbf{e}] +\frac{1}{2} \boldsymbol{\Omega}^* \wedge \mathbf{e}.\label{eq:e_dot}
\end{equation}
\end{subequations}
Here, $\boldsymbol{\Omega}^*=\bnabla^* \wedge \mathbf{u}^*$ is the vorticity, and $\beta^*$ the gyrotactic time scale. The random rotational motion is represented by a constant isotropic diffusivity $D_R^*$ \citep{Pedley2010b}. We also note here that the cell is assumed to be spherical in (\ref{eq:e_dot}), so that the effect of cell eccentricity (hence the effect of strain rate) is neglected. This assumption excludes the appearance of instability mechanisms due to the cell shape: for example, rod-like swimming cells in suspension may yield the instability proposed by \cite{Koch1989} and \cite{Saintillan2006}. However, for more rounded algal species like those in the \textit{Chlamydomonas} genus, \cite{OMalley2012} showed that the effective cell eccentricity of a swimming cell is much smaller than that of an inanimate body due to the influence of flagella beating. Therefore, in effect, a spherical approximation of the cell is a reasonable approximation at the continuum level.

In model F, the average cell swimming direction is obtained as
\begin{equation}\label{eq:eavg}
\langle \mathbf{e} \rangle \equiv \int_{\|\mathbf{e}\|=1} ~\mathbf{e} f(\mathbf{e})~d^2\mathbf{e}.
\end{equation}
The diffusivity tensor is given by the covariance matrix multiplied by a correlation time scale $\tau^*$: i.e.
\begin{equation}\label{eq:fk_diff}
{\diffusion}_F^*={V_c^*}^2 \tau^*
(\langle \mathbf{e} \mathbf{e}\rangle -\langle \mathbf{e} \rangle \langle \mathbf{e} \rangle),
\end{equation}
where the subscript $F$ in place of $m$ denotes the diffusivity obtained from model F. As discussed by \cite{Pedley2010b}, such a form of diffusivity is essentially \emph{ad hoc}, as the correlation time scale takes into account both the natural variability of the cell properties in the population and the randomness in the cell swimming direction. Therefore, the correlation time scale $\tau^*$ has often been used from the values directly obtained from experimental measurements \citep{Hill1997,Vladimirov2000}. In the present study, we will use $\tau^*=5.35 s$, which is slightly different from $\tau^*=5 s$ used in previous studies \cite[e.g.][]{HP2014b}, as this will allow us to set model F and G to share the same radial diffusivity in stationary suspension. A detailed discussion on this issue will be given in \S \ref{sec:Parameters}. 

\subsubsection{Model G}
Model G was developed from the Smoluchowski equation, which describes the probability density function in both physical and orientational space.
In this respect, the `diffusivity' in (\ref{eq:dim_n_gen}) would be interpreted as a consequence of the dispersion resulting from the interaction between the random cell orientation and the translational swimming motion. 
Model G uses the averaged cell swimming direction vector same as that of model F (i.e. (\ref{eq:eavg})). However, it gives a different form for the effective diffusivity \cite[]{Manela2003,Hill2002}:
\begin{equation}
    {\diffusion}_G^*=\int_{\|\mathbf{e}\|=1}  \left[ f(\mathbf{e})\mathbf{B}^*(\mathbf{e})\mathbf{e} {V^*_c}+f(\mathbf{e})\mathbf{B}^*(\mathbf{e})\mathbf{B}^*(\mathbf{e}) \cdot \mathsfbi{G}^* \right]^{sym} d^2 \mathbf{e}, \label{eq:GTD_diff}
\end{equation}
where the subscript $G$ in the place of the subscript $m$ indicates that the diffusivity originates from model G, and $\mathsfbi{G}^*(\equiv\nabla^*\textbf{u}^*$) is the local velocity gradient tensor. In (\ref{eq:GTD_diff}), $f(\mathbf{e})$ is obtained from (\ref{eq:pdf_gen}) as in model F, and the vector $\mathbf{B}^*(\mathbf{e})$ satisfies 
\begin{subequations} \label{eq:GTD_B}
\begin{equation}
    \bnabla_e \cdot \left[ \mathbf{\dot{e}^*}f\mathbf{B}^*- D_R^* \bnabla_e(f\mathbf{B}^*)\right]-f\mathbf{B}^* \cdot \mathsfbi{G}^* = ~V_c^*( \mathbf{e}-\eavg)f,
\end{equation}
subject to
\begin{equation}
     \int_{\|\mathbf{e}\|=1} f \mathbf{B}^*(\mathbf{e}) d^2\mathbf{e}=\mathbf{0}.
\end{equation}
\end{subequations}
Here, $\mathbf{B}^*(\mathbf{e})$ is the long-time limit of the difference between the overall average position, and the average position of the cell given its instantaneous orientation is $\mathbf{e}$ \citep{Hill2002}.
Thus, the form of (\ref{eq:GTD_B}) implies that the dimension of $\mathbf{B}^*$ is length, and its appropriate scale would be $V_c^* /D_R^*$.

\subsection{Non-dimensionalisation and boundary conditions}\label{sec:nondim}
The equations of motion (\ref{eq:dim_contin}), (\ref{eq:dim_vel}), (\ref{eq:dim_n_gen}), and (\ref{eq:pdf_gen}) are non-dimensionalised by 
\refstepcounter{equation}
$$
  \mathbf{x}=\frac{\mathbf{x}^*}{h^*}, \quad
  t=\frac{t^* V_c^*}{h^*}, \quad
  \mathbf{u}=\frac{\mathbf{u}^*}{V_c^*}, \quad
  p=\frac{p^*}{\rho^*V_c^{*2}}, \quad
  n=\frac{n^*}{N^*}, 
  \eqno{(\theequation{\mathit{a}-\mathit{e}})} \label{eq:non_dimensionalise}
$$
where $N^*$ is the cell concentration if the suspension is set to be uniform, and $h^*$ is the pipe radius. The resulting dimensionless equations are 
\begin{subequations}\label{eq:non_dim}
\begin{eqnarray}
  \bnabla \bcdot \mathbf{u}  & = & 0, \label{eq:contin} \\
  \pardt{\mathbf{u}} + (\mathbf{u} \bcdot \bnabla) \mathbf{u} & = & -\bnabla p + \frac{1}{\Rey} \nabla^{2} \mathbf{u} + \Ri~n~ \mathbf{k},  \label{eq:vel} \\
  \pardt{n} + \bnabla \bcdot [(\mathbf{u} + \eavg ) n] & = & \frac{1}{D_R} \bnabla \bcdot ({\diffusion}_m \bcdot \bnabla n), \label{eq:n}  \\
  \bnabla_e \bcdot \left[ \lambda[-\mathbf{k}+(\mathbf{k} \bcdot \mathbf{e})\mathbf{e}]f+\frac{1}{2 D_R} \boldsymbol{\Omega} \wedge \mathbf{e} f \right]  & = &   \bnabla^2_e f, \label{eq:pdf}
\end{eqnarray}
with boundary conditions
    \begin{equation} \label{eq:vel_BC}
        \mathbf{u} |_{r=1}  =  (0,0,0)^T,
    \end{equation}
    \begin{equation} \label{eq:n_BC}
      \left[(\mathbf{u} +  \eavg ) n  - \frac{1}{D_R} {\diffusion}_m \bcdot \bnabla n \right] |_{r=1} \bcdot \mathbf{i}  = 0,
    \end{equation}
where
\begin{eqnarray}
\Ri=\ddfrac{N^* \upsilon^* g'^*h^*}{V_c^{*2}}, &
\Rey=\ddfrac{V_c^*h^*}{\nu^*}, &
\lambda=\ddfrac{1}{2\beta^* D_R^*}, ~
D_R=\frac{D_R^* h^*}{V_c^*}.\label{eq:non_dim_para}
\end{eqnarray}
Here, $\Rey$ is the Reynolds number based on the cell swimming speed, $\Ri$ the Richardson number, and $D_R$ the dimensionless rotational diffusivity. We note that (\ref{eq:pdf}) is further divided by $D_R$, since the appropriate time scale in the cell-orientation space would be $1/D_R^*$, different from $h^*/V_c^*$ in the physical space. 
Lastly, the dimensionless translational diffusivities for model F and G are given by  
\begin{equation}
   \diffusion_F=\ddfrac{\diffusion^*_F D_R^*}{V_c^{*2}}=\tau (\langle \mathbf{e} \mathbf{e}\rangle -\langle \mathbf{e} \rangle \langle \mathbf{e} \rangle) \label{eq:FK_Diff_nondim}
\end{equation} 
and 
\begin{equation}
   \diffusion_G=\ddfrac{\diffusion^*_G D_R^*}{V_c^{*2}}=\int_{\|\mathbf{e}\|=1}  \left[ f(\mathbf{e})\mathbf{B}(\mathbf{e})\mathbf{e} +f(\mathbf{e})\mathbf{B}(\mathbf{e})\mathbf{B}(\mathbf{e}) \cdot \mathsfbi{G} \right]^{sym} d^2 \mathbf{e}, \label{eq:GTD_Diff_nondim}
\end{equation}
respectively, where $\tau=\tau^* D_R^*$, $\mathbf{B}=\mathbf{B}^* D_R^*/V_C^*$ and $\mathsfbi{G}=\mathsfbi{G}^*/D_R^*=\bnabla \mathbf{u}/D_R$. As mentioned previously, $\tau$ in (\ref{eq:FK_Diff_nondim}) will be chosen appropriately later. To compute $\diffusion_G$, it is convenient to introduce $\mathbf{b}(\mathbf{e})=f(\mathbf{e})\mathbf{B}(\mathbf{e})$, which from (\ref{eq:GTD_Diff_nondim}) satisfies the following equation:
\begin{equation} \label{eq:GTD_b}
    \bnabla_e \cdot \left[ \mathbf{\dot{e}}\mathbf{b}- \bnabla_e\mathbf{b}\right]-\mathbf{b} \cdot \mathsfbi{G} = ~(\mathbf{e}-\eavg)f,
\end{equation}
where $\mathbf{\dot{e}}= \mathbf{\dot{e}^*} / D_R^*$. Once $\mathbf{b}(\mathbf{e})$ is obtained from (\ref{eq:GTD_b}), the translational diffusivity of model G is given by the following expression:
\begin{equation}
   \diffusion_G=\int_{\|\mathbf{e}\|=1}  \left[ \mathbf{b}(\mathbf{e})\mathbf{e} +\frac{\mathbf{b}(\mathbf{e})\mathbf{b}(\mathbf{e})}{f(\mathbf{e})} \cdot \mathsfbi{G} \right]^{sym} d^2 \mathbf{e}. \label{eq:GTD_Diff_nondim_b}
\end{equation}
\end{subequations}

\subsection{Basic state}
The basic state of (\ref{eq:non_dim}) is first calculated by assuming that the velocity and cell-concentration fields are steady, axisymmetric and homogeneous along the axial direction: i.e.
\refstepcounter{equation}
$$
\mathbf{u}=\mathbf{u_0}=(0,0,U(r)), \qquad n=N(r), \qquad  \pardt{}=\pardz{}=\pardpsi{}=0. 
\eqno{(\theequation{\mathit{a}-\mathit{c}})}
$$
The basic-state pressure $P_0(r,z)$ is obtained by integrating (\ref{eq:vel}) in the radial direction:
\begin{subequations}
\begin{equation}
 \pardz{P_0}=\frac{2}{Re} \pardr{U}|_{r=1} +Ri, \label{eq:total_pressure}
\end{equation}
which is composed of the pressure gradient driving the flow,
\begin{equation}
 \pardz{P_0^d}=\frac{2}{Re} \pardr{U}\Big|_{r=1},
\end{equation}
\end{subequations}
and the hydrostatic pressure gradient balancing out the gravitational term (i.e. $Ri$) in (\ref{eq:total_pressure}).

Removal of the hydrostatic balance from (\ref{eq:non_dim}) then yields the following equations for basic state:
\begin{subequations} \label{eq:base_full}
    \begin{equation}\label{eq:U0}
        - r\pardz{P^d_0}+\frac{1}{Re}\pardr{} r \pardr{U}+r Ri (N-1)=0,
    \end{equation}
    \begin{equation}\label{eq:v0}
        - \pardr{P^d_0}=0,
    \end{equation}
    \begin{equation}\label{eq:n0}
        \pardr{}(r N \esth{r}_0)=\frac{1}{D_R}\pardr{}(r {D}_{rr,0} \pardr{N}),
    \end{equation}
with boundary conditions
    \begin{equation} \label{eq:vel0_n0_BC}
        U(1) = 0, \quad  \left[\esth{r}_0 N-\frac{{D}_{rr,0}}{D_R}\pardr{N} \right] \Big|_{r=1} = 0,
    \end{equation}
and the compatibility condition at the centre
    \begin{equation}\label{eq:vel0_n0_BC_axis}
       \left. \pardr{U} \right|_{r=0} = 0, \left. \quad \pardr{N} \right|_{r=0} = 0.
    \end{equation}
\end{subequations}
Here, the subscript $0$ in $\esth{r}_0$ and ${D}_{rr,0}$ indicates the variables from \S\ref{sec:nondim}, which are used for the calculation of steady basic state. 

Since the total number of the cells is preserved over the given control volume, we impose the normalisation condition for the cell concentration,
\begin{equation}\label{eq:n_norm}
    \int_0^1 N(r) r dr=\frac{1}{2}.
\end{equation}
The flow rate $Q$ is assumed to be given, such that
\begin{equation}\label{eq:Q}
    \int_0^1 U(r) r dr=\frac{Q}{2 \pi}.
\end{equation}

\subsection{Linear stability analysis}
Now, we consider a small perturbation about the basic state:
\begin{equation}
        \mathbf{u}=\mathbf{u_0}+\epsilon \mathbf{u'}+\textit{O} (\epsilon^2), \quad p=P_0+\epsilon  p'+\textit{O} (\epsilon^2),  \quad n=N+\epsilon n'+\textit{O} (\epsilon^2),
\end{equation}
where $\mathbf{u_0}=(0,0,U(r))$, $\mathbf{u'}=(v',w',u')$ and $\epsilon \ll 1$. We also define $\boldsymbol{\Omega_0}=\nabla \wedge \mathbf{u_0}$, $\boldsymbol{\Omega}'=\nabla \wedge \mathbf{u}'$, $\mathsfbi{G_0}=\nabla \mathbf{u_0}/D_R$ and $\mathsfbi{G}'=\nabla \mathbf{u}'/D_R$ accordingly. Here, we note that the axial velocity perturbation and the cell concentration over a given control volume $V_0$ should satisfy
\begin{equation}
   \int_{V_0} n' dV= \int_{V_0} u' dV=0,
\end{equation}
as the flow rate $Q$ is fixed and the total number of the cells over the entire domain is preserved in time. The linearised equations for the small perturbation are then given as
\begin{subequations}\label{eq:lin_full}
    \begin{equation}
        \pardz{u'}+\frac{1}{r}\pardr{rv'}+\frac{1}{r}\pardpsi{w'}=0,
    \end{equation}
    \begin{equation}
        \pardt{u'}+U \pardz{u'}+v' \pardr{U}=-\pardz{p'}+\frac{1}{Re}\nabla^2 u' + \Ri~ n',
    \end{equation}
    \begin{equation}
        \pardt{v'}+U \pardz{v'}            =-\pardr{p'}+\frac{1}{\Rey}(\nabla^2 v' - \frac{v'}{r^2} - \frac{2}{r^2}\pardpsi{w'}),
    \end{equation}
    \begin{equation}
        \pardt{w'}+U \pardz{w'}            =-\frac{1}{r}\pardpsi{p'}+
        \frac{1}{\Rey}(\nabla^2 w' - \frac{w'}{r^2} +\frac{2}{r^2}\pardpsi{v'}),
    \end{equation}
    \begin{eqnarray}\label{eq:n_pert}
        \pardt{n'} &+& 
            n'(\frac{\eo{r}}{r}+\pardr{\eo{r}})+\eo{r}\pardr{n'} +U \pardz{n'}+\eo{z} \pardz{n'} +\frac{\eo{z}}{r} \pardpsi{n'} + v' \pardr{N} \nonumber \\
        & + & 
            {\esth{r}'}\pardr{N} + N \pardr{{\esth{r}'}}+N \frac{\esth{r}'}{r}+N \pardz{\esth{z}'}+\frac{N}{r}\pardpsi{\esth{\psi}'} \nonumber \\
        & = &
            \frac{1}{D_R}\left[ \frac{D_{rr}'}{r}\pardr{N}+\pardr{D_{rr}'}\pardr{N}+D_{rr}' \pardrr{N}+\pardz{D_{rz}'}\pardr{N} +\frac{1}{r}\pardz{D_{r\psi}'}\pardr{N}\right. \nonumber \\
        & + &
            \frac{1}{r} \left(D_{rr,0}\pardr{n'}+D_{rz,0}\pardz{n'}+2D_{r\psi,0}\frac{\partial^2 n'}{\partial r \partial \psi}+\pardr{D_{r\psi,0}}\pardpsi{n'}+2D_{\psi z,0}\frac{\partial ^2 n'}{\partial \psi \partial z} \right)  \nonumber \\
        & + &
            \pardr{D_{rr,0}}\pardr{n'}+2\pardr{D_{rz,0}}\pardz{n'}+D_{rz,0}\frac{\partial ^2 n'}{\partial r \partial z} \nonumber \\
        & + &
            \left. D_{rr,0}\pardrr{n'}+D_{zz,0}\pardzz{n'}+\frac{1}{r^2}D_{\psi \psi,0}\pardpsipsi{n'}
             \right],
    \end{eqnarray}
with the boundary conditions at the wall
    \begin{equation}\label{eq:velp_BC}
        u'|_{r=1} =v'|_{r=1}=w'|_{r=1}= 0,
    \end{equation}
    \begin{equation}\label{eq:np_BC}
      N v'+ N \esth{r}' + n' \eo{r}=\frac{1}{D_R}
      \left(D_{rr}' \pardr{N}+D_{rr,0} \pardr{n'}+D_{rz,0}\pardz{n'}+\frac{D_{r\psi,0}}{r}\pardpsi{n'} \right).
    \end{equation}
\end{subequations}
Here, $\diffusion'_m$ and $\eavg'$ are computed by linearising (\ref{eq:pdf}) and (\ref{eq:GTD_b}) in a similar manner to \cite{HP2014a,HP2014b}. The detailed set of lengthy linearised equations can be found in Appendix \ref{sec:lin_diff}.

Finally, the following normal-mode solution is introduced for linear stability analysis:
\begin{eqnarray}\label{eq:normal_mode}
&& \mathbf{u}'(r,\psi,z,t)=\hat{\mathbf{u}}(r)e^{i(\alpha z+m \psi-\omega t)}+c.c,
~p'(r,\psi,z,t)=\hat{p}(r)e^{i(\alpha z+m \psi-\omega t)}+c.c. \nonumber \\
&& n'(r,\psi,z,t)=\hat{n}(r)e^{i(\alpha z+m \psi-\omega t)}+c.c,
\end{eqnarray}
where $\alpha$ is the real axial wavenumber, $m$ the wavenumber in the azithumal direction (positive integer), and $\omega$ the complex frequency. Substitution of the normal-mode solution into (\ref{eq:lin_full}) yields an eigenvalue problem, as detailed in Appendix \ref{sec:lin_modal}. 

\subsection{Numerical methods\label{sec:numerical}}
To obtain the basic state from (\ref{eq:base_full}) and subsequently examine its stability, $\eavg$ and $\diffusion_m$ in (\ref{eq:n}) need to be computed as a function of $\boldsymbol{\Omega}/D_R$ (model F and G) and $\mathsfbi{G}$ (model G): see \S\ref{subsec:GTD_FK}. The steady solution $f(\mathbf{e})$ to (\ref{eq:pdf}) is obtained using the solver identical to that of \cite{HP2014a}, and the reader may refer to \S 3  in \cite{HP2014a} for the detailed result on $\eavg$ and $\diffusion_F$. As for model G, the mean swimming orientation vector $\eavg$ is the same as the one for model F. Therefore, only $\diffusion_G$ used for model G needs to be obtained. With the same discretisation schemes and resolution used for (\ref{eq:pdf}) in \cite{HP2014a},  (\ref{eq:GTD_b}-\ref{eq:GTD_Diff_nondim_b}) are solved to obtain $\diffusion_G$. The numerical results agree very well with the ones given by \cite{Bearon2012}. Similarly, $\eavg'$ and $\diffusion_m'$ used in (\ref{eq:n_pert}) are calculated by applying the same discretisation schemes to the equations in Appendix \ref{sec:lin_diff}. 

In order to compute the basic state, the radial direction of (\ref{eq:base_full}) is discretised using a Chebyshev collocation method described in \cite{Weideman2000}. Two solvers have been written in \texttt{MATLAB}: one is based on a Newton-Ralphson method, and the other solves the unsteady version of  (\ref{eq:base_full}) to obtain its steady solution, the latter of which is used to study the stability of the basic state to axially uniform and radially independent perturbation. The solutions from the two solvers have been validated against each other when stable solutions are admitted. They are found to be identical up to the given numerical precision.  Finally, for the purpose of studying bifurcation of the basic state, the Newton-Raphson solver is combined with a pseudo-arclength continuation algorithm (see \S\ref{sec:Results_basic_state}). 

For the linear stability, the radial direction of (\ref{eq:lin_full_modal}) is discretised using the same discretisation method as the one for the basic state. The discretised eigenvalue problem is solved using the function \texttt{eig} in \texttt{MATLAB}. The computation is performed with $N_r=175$, showing no difference from the results with $N_r=100$. The computed eigenvalues for $\mathrm{Ri}=0$ are validated against pipe flow stability data of \cite{Schmid1994,Schmid2001} and \cite{Meseguer2003} with excellent agreement.

\subsection{Parameters\label{sec:Parameters}}
\begin{table}
  \begin{center}
\def~{\hphantom{0}}
  \begin{tabular}{lccc}
   Parameter    & Description   &   Reference Value &             Units    \\ [3pt]
   \hline 
   $~\rho^*~$     & Fluid density &   $~1~$ &               $~\mathrm{g/cm^{3}}~$ \\
   $~g^*~$        & Gravitational acceleration &   $~980~$ &               $~\mathrm{cm/s^2}~$ \\
   $~\nu^*~$        & Dynamic viscosity &   $~0.00957~$ &               $~\mathrm{cm^{2}/s}~$ \\
   $~h^*~$     & Radius of pipe &   $~0.1 \sim 0.4~ (0.19)$ &               $~\mathrm{cm}~$ \\
   $~N^*$     & Average cell number density &   $0 \sim 3.13 \times 10^{5}~$ &   $~\mathrm{cells/cm^3}~$ \\
   $~\Delta \rho/\rho~$  & Relative cell concentration &   $~0.05~$ &               $~-~$ \\
   $~\upsilon^*~$ & Cell volume & $~2.1\times 10^{-9}~$    &          $~\mathrm{cm^3}~$  \\
   $~g'(=g^*\Delta \rho/\rho) ~$ & Relative gravity & $~49~$    &   $~\mathrm{cm/s^{2}}~$  \\
   $~\beta^*~$ & gyrotactic time scale  & $~3.4~$   & $~\mathrm{sec}~$  \\
   $~V_c^*~$ & Swimming speed & $~6.3 \times 10^{-3}~$  &  $~\mathrm{cm/s}~$  \\
   $~\tau^*~$ & Correlation time scale & $~5.35~$   &  $~\mathrm{s}~$  \\
   $~D_V^*(={V_c^*}^2\tau^*)~$ & Nominal translation cell diffusivity & $~2.12 \times 10^{-4}~$   &  $~\mathrm{cm^2/s}~$  \\
   $~D_R^*~$ & Rotational diffusivity & $~0.067~$   &  $~\mathrm{1/s}~$  \\
  \end{tabular}
  \caption{Parameters and their reference values in the present study. Most of the parameters for the cell properties are taken from the data for \emph{C. Nivalis} \cite[]{Pedley1990,Bees1998,Pedley2010b}. }
  \label{tab1}
  \end{center}
\end{table}

\begin{table}
  \begin{center}
\def~{\hphantom{0}}
  \begin{tabular}{lcc}
   Parameter    & Description   &   Reference Value              \\ [3pt]
   \hline 
   $~\mathrm{Ri}~$     & Richardson number &   $0 \sim 160$  \\
   $~Q~$             & Dimensionless Flow Rate &   $0 \sim 7$  \\
   $~\mathrm{Re}~$        & Reynolds number based on $V_c^*$ &   $0.066 \sim 0.26~(0.126)$  \\
   $~\lambda~$     & gyrotactic time scale divided by $D_R^*$ &   $~2.2~$  \\
   $~D_R~$        & Rotational diffusivity normalised by $V_c^*/h^*$ &   $~2.13~$  \\
  \end{tabular}
  \caption{Dimensionless parameters and their values in the present study.}
  \label{tab2}
  \end{center}
\end{table}

A list of parameters and their values used in the present study are summarised in table \ref{tab1}. Here, it should be noted that $\tau^*$ is not a free parameter, as mentioned in \S \ref{subsec:GTD_FK}. Instead,  $\tau^*$ is set, such that model F and G provide consistent and quantitatively comparable results. Therefore, in the present study, we follow the approach of \cite{Bearon2012}, in which $\tau^*$ is chosen to match $D_{G,rr}$ with $D_{F,rr}$ in the absence of any flow (i.e. stationary medium). This gives $\tau=0.36$, consistent with that of \cite{Bearon2012}.

In this study, the radius of the cylindrical pipe is chosen to be around $0.2$cm, so that we obtain results comparable with the same as \cite{HP2014b}, at around $0.2$cm. The parameters for cell's biological properties, including cell swimming speed $V_c^*$, gyrotactic time scale $\beta^*$, relative cell density $\Delta \rho / \rho$ and rotational diffusivity $D_R^*$, are taken for \emph{C. nivalis} from previous studies \cite[e.g.][]{Pedley1990,Bees1998,Pedley2010b,HP2014b}, except the shape of the cell which we assume to be spherical. As in the linear stability analysis of \cite{HP2014b} and the experiment of \cite{Croze2017}, only the flow rate $Q$ and the background cell concentration (represented by $Ri$) are varied in this study. Based on the parameter values given in table \ref{tab1}, the dimensionless parameters and their values examined are given in table \ref{tab2}.


\section{Basic state \label{sec:Results_basic_state}}

\subsection{Very low flow rate \label{sec:vlowQ}}
\begin{figure}
	\sidesubfloat[]{\includegraphics[width = 0.45 \columnwidth]{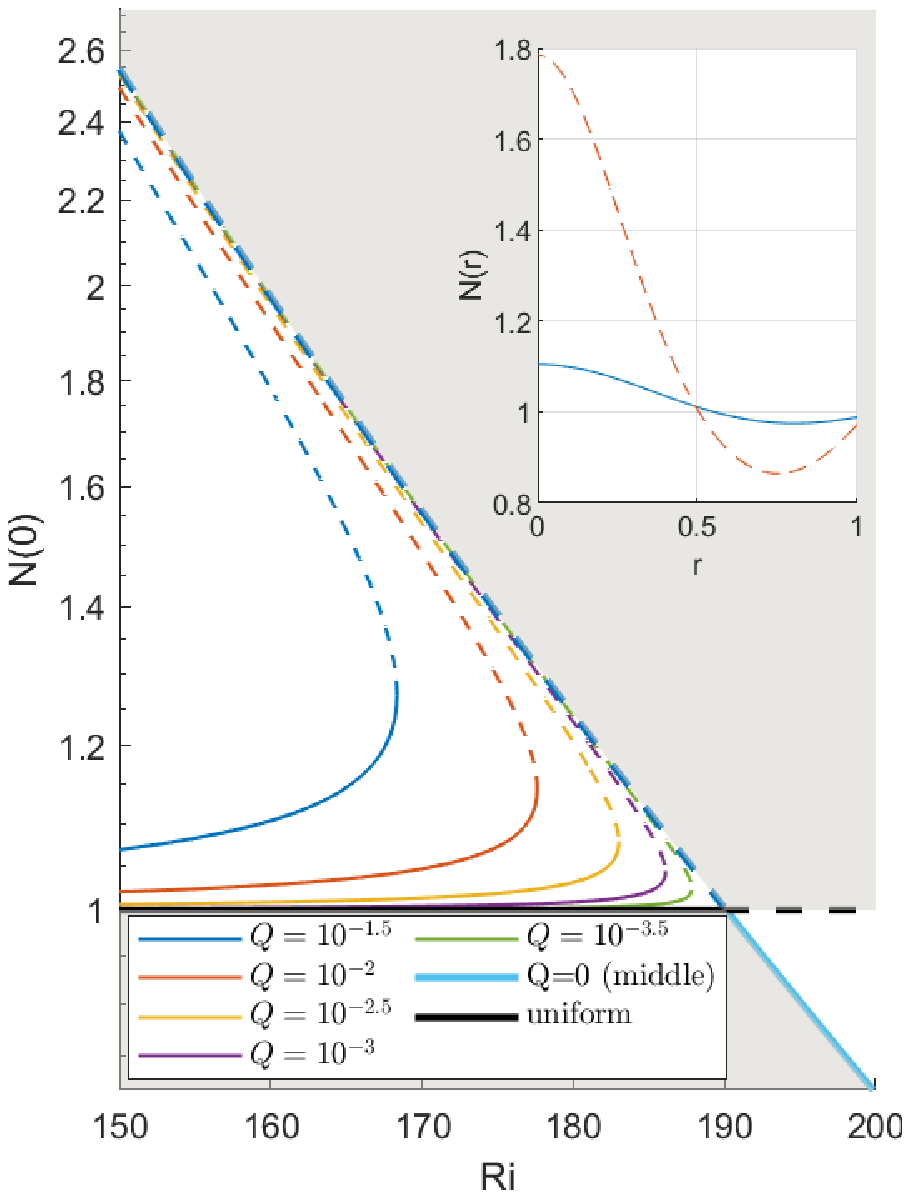}}
	\sidesubfloat[]{\includegraphics[width = 0.45 \columnwidth]{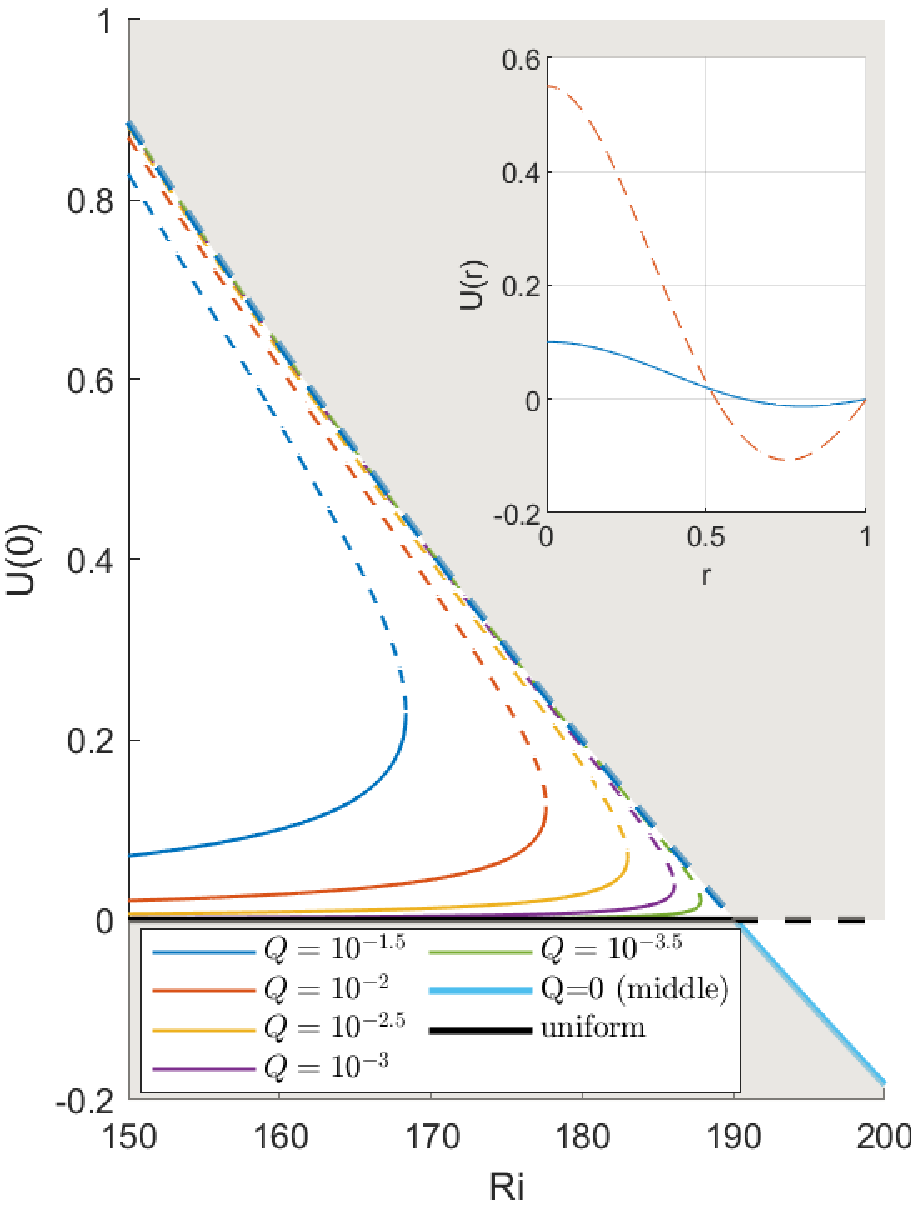}}
    \caption{Bifurcation of steady solutions with $Ri$ at very low flow rate \deleted{(model G)}: $(a)$ the cell concentration $N(0)$ and $(b)$ the centreline velocity $U(0)$ along the pipe axis. Here, \protect \solid, stable; \protect \dashed, unstable. \added{The grey area indicates where $Q<0$ contours would be. The insets show $N(r)$ and $U(r)$ at $Ri=160$: blue \protect \solid, the lower branch; red \protect \dashed, the middle branch. Note that as $Q \rightarrow 0$, the results from model F and G coincides.}}
    \label{fig:2}
\end{figure}

At very low flow rate (i.e. $Q \rightarrow 0$), model F and G share the same asymptotic value for $\eavg$ and the radial diagonal component of $\diffusion_m$ (see \cite{Bearon2012} and figure \ref{fig:7}). Therefore, the two models would give almost identical result when $Q$ is small enough. Hence, here we report the result from model G first in this case. 

When $Q=0$, a stationary uniform suspension (i.e. $U(r)=0$ and $N(r)=1$) is a solution to (\ref{eq:base_full}). However, a further numerical search has also found that there exists another solution at $Q=0$ featured with non-trivial $U(r)$ and $N(r)$, consistent with \cite{Bees2010a}. Bifurcation diagrams of these two solutions with respect to $Ri$ are shown for small positive $Q$ in figure \ref{fig:2}, where the centreline concentration $N(0)$ and axial velocity $U(0)$ are used to represent the state of the steady solutions. For $Q=0$, the two solutions meet at $Ri= Ri_c(\simeq 189.9)$, and their stability has been checked using the unsteady solver described in \S\ref{sec:numerical}. The stationary solution ($U(r)=0$ and $N(r)=1$) is found to be stable for $Ri<Ri_c$, but becomes unstable for $Ri>Ri_c$. On the other hand, the non-trivial solution, featured with a downflow and a high cell concentration along the pipe axis, is unstable for $Ri<Ri_c$. This solution gains its stability when $Ri>Ri_c$, and the form of the solution is subsequently changed with an upflow and a lower cell concentration along the pipe axis. The interchange of the stability of the two solutions at $Ri=Ri_c$ indicates that the stationary suspension in the vertical pipe experiences a transcritical bifurcation with $Ri$.

When a small downflow ($Q>0$) is applied, the transcritical bifurcation point given at zero flow rate turns into a saddle-node point. \replaced{At $Ri>Ri_c$, the axial velocity is upward instead of downward (i.e. $U(0)<0$), even though the overall net flow is downward ($Q>0$). If a small upflow ($Q<0$) is applied instead, then the opposite is true, as shown by the grey area in figure \ref{fig:2}. }{A similar change of the bifurcation is also observed when a upflow ($Q<0$) is applied (not shown here). }The transition from a transcritical to a saddle-node bifurcation is a consequence of an imperfect bifurcation caused by the addition of a small non-zero flow rate. This behaviour of the solutions with two parameters, $Ri$ and $Q$, can be understood within the framework of co-dimension two bifurcation (i.e. bifurcation with two parameters). The overall bifurcation is also closely related to how the system would evolve with the flow direction. However, given the scope of the present study, we shall only focus on the downflowing case and leave the upflowing case as future work. 

\begin{figure}
\centering{}
	\sidesubfloat[]{\includegraphics[width = 0.512 \columnwidth]{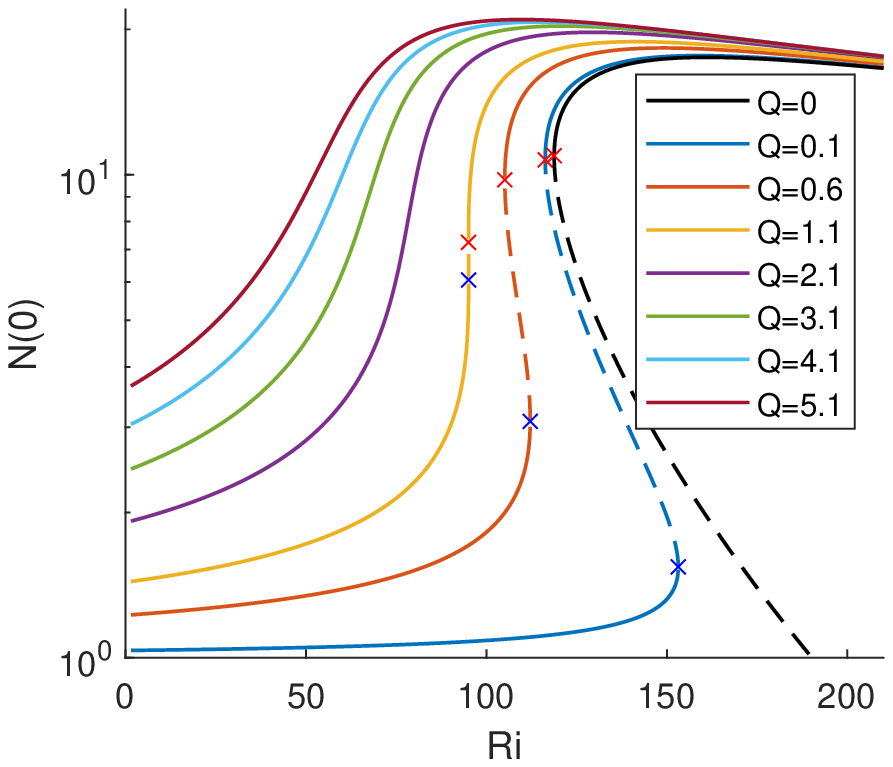}\label{fig:3a}}
	\sidesubfloat[]{\includegraphics[width = 0.387 \columnwidth]{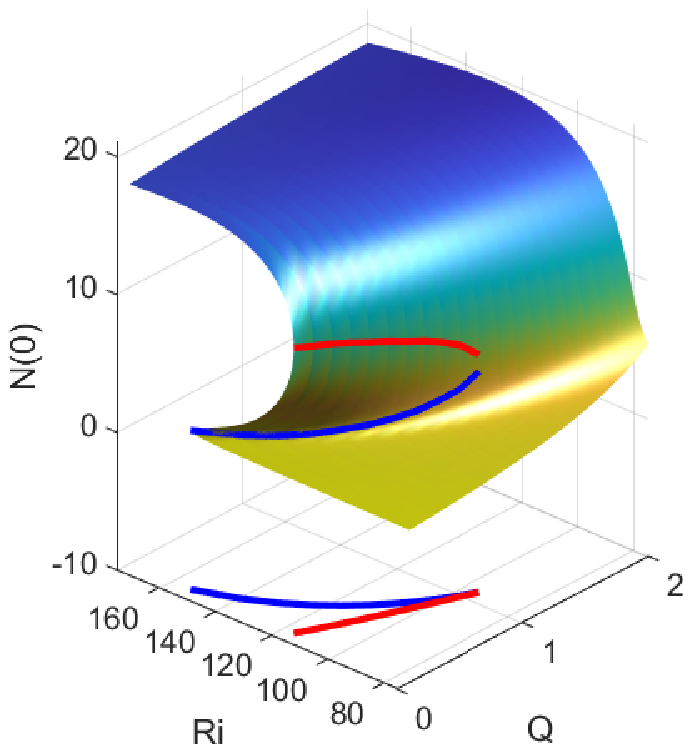}\label{fig:3b}}
\caption{Bifurcation of the steady solutions at high flow rates (model F): $(a)$ $N(0)$ with respect to $Ri$ for several flow rates $Q$; $(b)$ surface plot of $N(0)$ in the $Q-Ri$ space. In $(a)$, \protect \solid, stable; \protect \dashed, unstable. The blue cross (x) indicates the first saddle-node point where lower and middle branches meet, while the red cross (x) indicates the second saddle-node point where middle and upper branches meet. In $(b)$, the blue and red curves indicate the trajectories of the two saddle-node points with change in $Q$, respectively. The graph at the bottom shows the projection of the trajectories onto the $Q-Ri$ plane.}
\end{figure}

\begin{figure}
    \centering
    \includegraphics[width=0.9 \columnwidth]{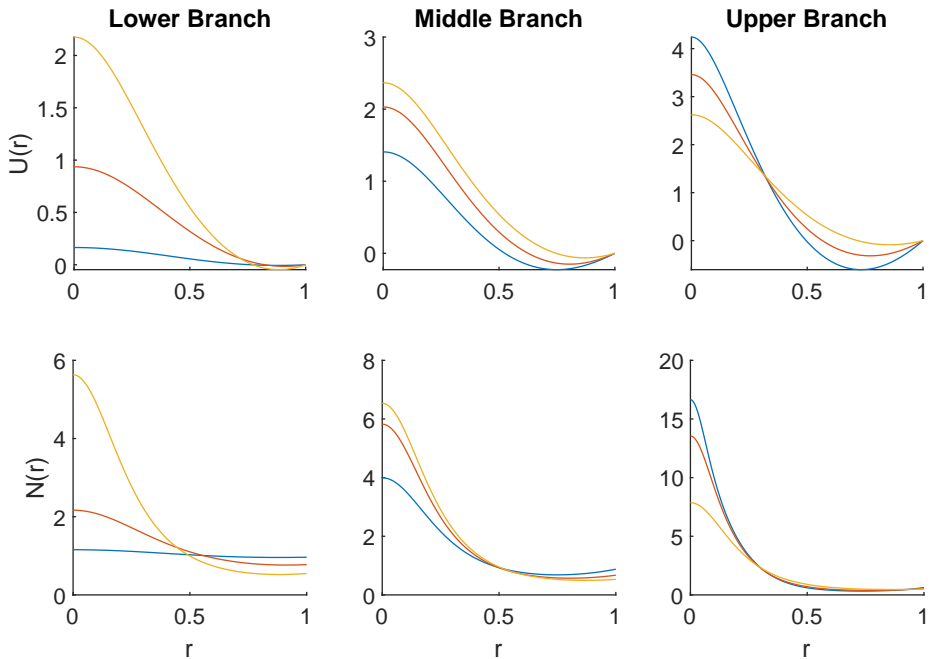}
    \caption{Downward velocity $U(r)$ (top) and cell concentration $N(r)$ (bottom) profiles of the steady basic states at \added{(blue) $Q=0.1;Ri=132$, (red) $Q=0.6;Ri=108$ and (yellow) $Q=1.1;Ri=95.04$} (model F). Here, the left, middle and right columns represent the lower-, middle- and upper-branch solutions, respectively. }
    \label{fig:4}
\end{figure}

\subsection{Model F}
The solution featured with $N(0)>1$ and $U(0)>0$ in figure \ref{fig:2} is further continued for much higher flow rates with model F. Figure \ref{fig:3a} shows the result from such continuation with model F, in which the concentration at the pipe axis $N(0)$ is used to represent the state of the corresponding steady solution. At sufficiently low flow rate ($Q\lesssim 1$), the bifurcation of the solutions featured with two saddle-node points (highlighted by the crosses in figure \ref{fig:3a}), indicating the emergence of three solutions at the given flow rate. With the increasing order of $N(0)$, the three types of solutions will be denoted by lower, middle and upper branches, respectively, and they are visualised in figure \ref{fig:4}. \added{Here, we note that $N(0)$ was found to uniquely represent each solution per parameter set, and hence was used as a representation for each solution.} In general, as the solution is continued from the lower to upper branch, both the velocity and cell-concentration profiles near the pipe axis ($r=0$) tend to be sharper. \added{This is expected because the lower-branch state is a homotopy of the uniform suspension obtained with increasing flow rate (i.e. state without gyrotactic instability), while the middle- and upper-branch states are homotopy of nonlinearly developed plumes from the gyrotatic instability at zero flow rate (i.e. state with nonlinearly staruated gyrotactic instability).} Examination of the stability of each solution to axially and radially uniform perturbations (i.e. unsteady calculation of the basic state) reveals that the lower and upper branches are stable, whereas the middle one is unstable, indicating bistable nature of the basic state featured with hysteresis. 

As the flow rate $Q$ is further increased, the two saddle-node points become closer. They eventually merge and disappear with the increasing $Q$. The curves in figure \ref{fig:3b} shows how the two saddle-node points evolved with $Q$ in the $Q-Ri-N(0)$ space. At the bottom of this figure, the trajectories of the two saddle-node points are projected onto the $Q-Ri$ plane to visualise how the $Ri$ values at these points change with $Q$. The two saddle-node points in the $Q-Ri$ plane indeed merge as $Q$ increases. Beyond this merging point, the bistable behaviour of the steady solutions disappears, as there exits only single steady solution. This type of codimension two bifurcation is a cusp bifurcation \citep[see][]{Zeeman1976,Thom1989}. Further discussion on the bifurcation in relation to previous experimental observations will be given in \S \ref{sec:Implication}.

Lastly, it is worth mentioning that, in model F, the density at the pipe axis $N(0)$ decreases with increasing $Ri$ for sufficiently large $Ri$ (see the upper branch in figure \ref{fig:3a}). As discussed in \cite{HP2014b}, the decrease of $N(0)$ is the direct consequence of increasing $D_{rr}$ with the increase of background base-flow shear (see figure \ref{fig:7b} of this paper and figure 3$(b)$ of \cite{HP2014b}), which smooths out the gyrotactic focusing near the pipe axis.

\subsection{Model G\label{subsec:Bifur_GTD_K}}
\begin{figure}[ht]
\centering{}
	\sidesubfloat[]{\includegraphics[width = 0.512 \columnwidth]{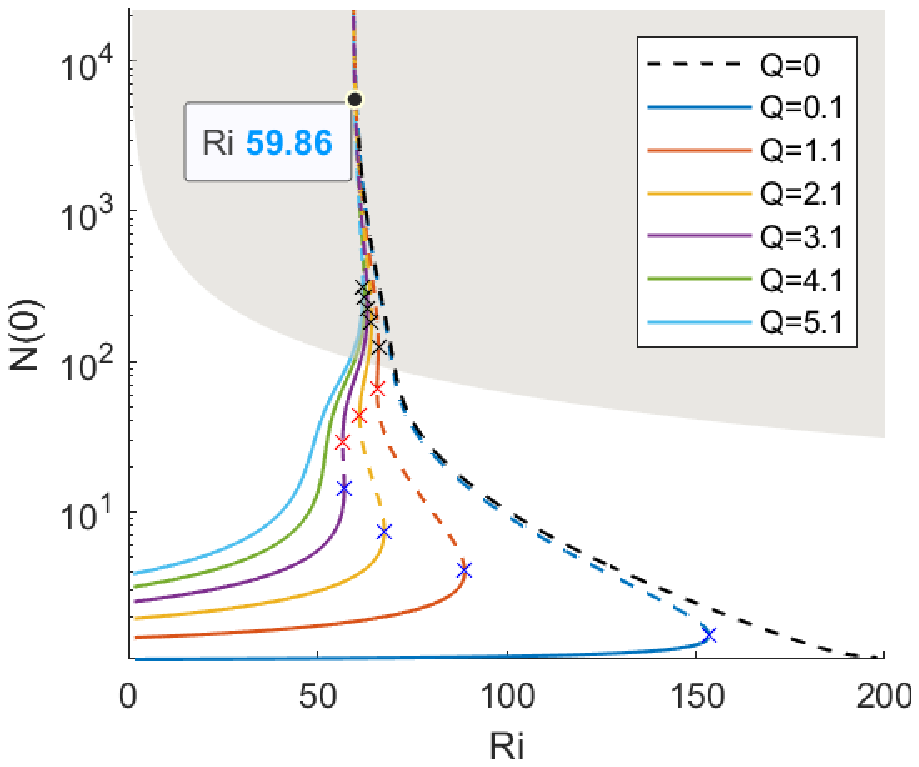}\label{fig:5a}}
	\sidesubfloat[]{\includegraphics[width = 0.387 \columnwidth]{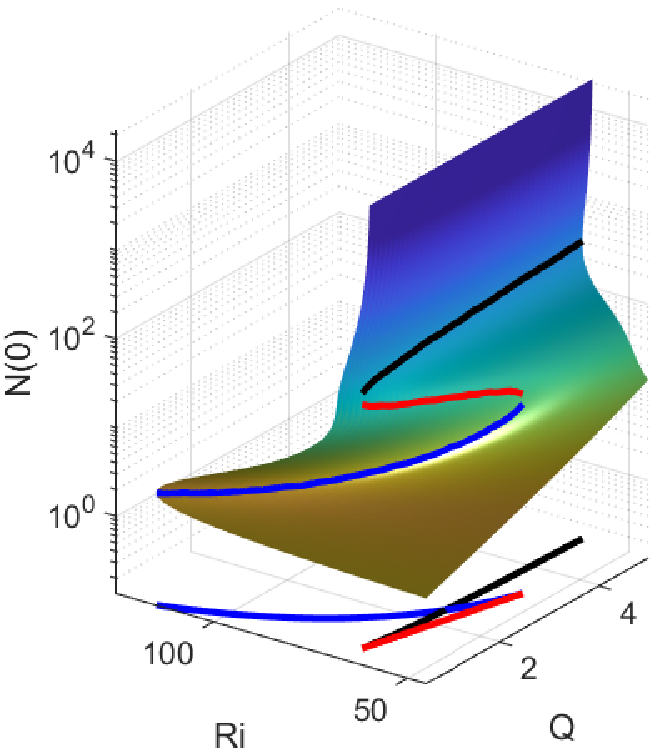}\label{fig:5b}} \\
	\sidesubfloat[]{\includegraphics[width = 0.3 \columnwidth]{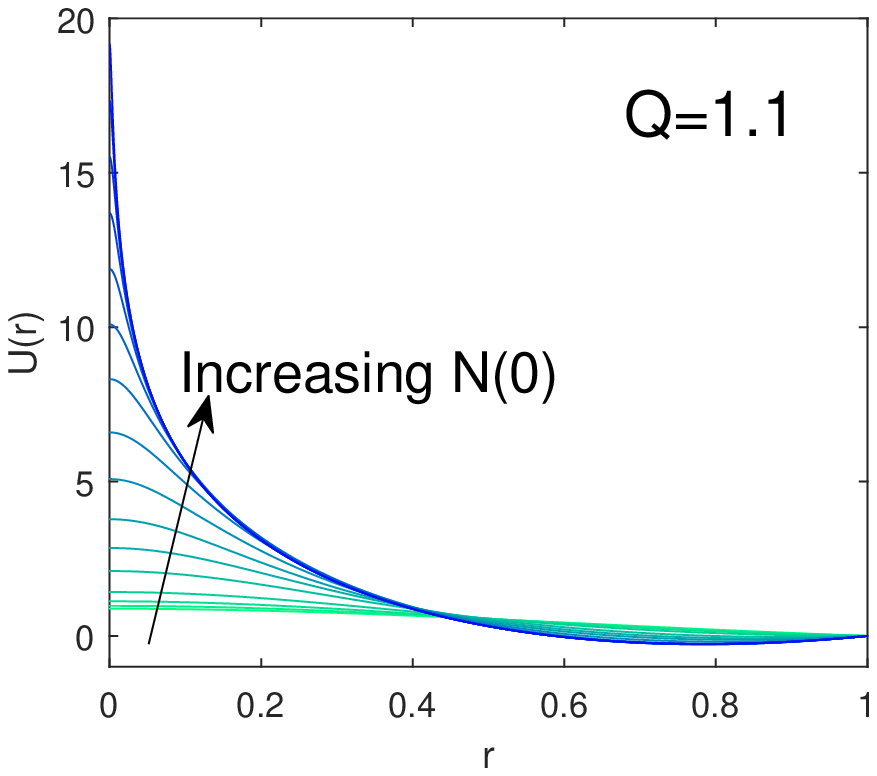}\label{fig:5c}}
	\sidesubfloat[]{\includegraphics[width = 0.3 \columnwidth]{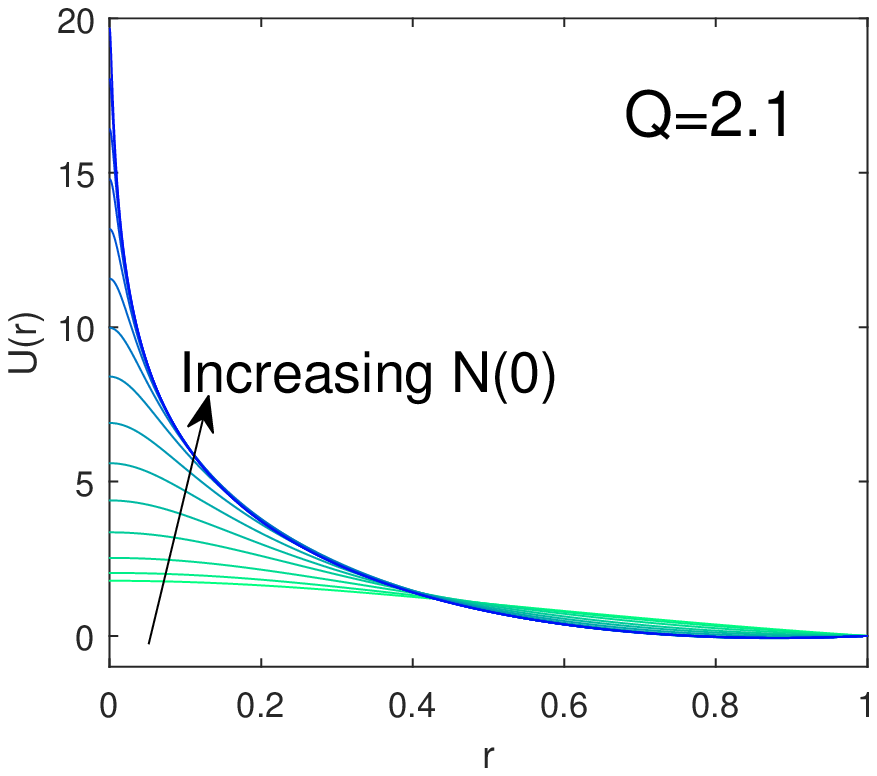}\label{fig:5d}}
	\sidesubfloat[]{\includegraphics[width = 0.3 \columnwidth]{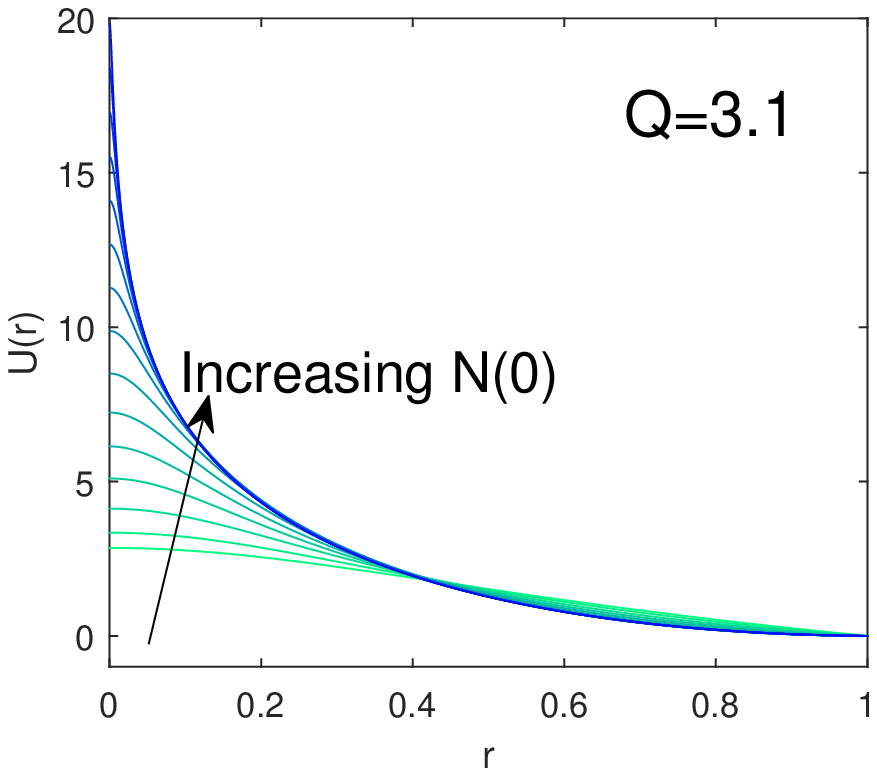}\label{fig:5e}}
\caption{Bifurcation of the steady solutions at high flow rates (model G): $(a)$ $N(0)$ with respect to $Ri$ for several flow rates $Q$; $(b)$ Surface plot of $N(0)$ in the $Q-Ri$ space. In $(a)$, \protect \solid, stable; \protect \dashed, unstable. Here, the blue, red and black crosses (x) indicate the first, second and third saddle-node points, respectively, as the solution is continued from the lower to the upper branch. The grey area indicates the cases where the volume fraction at the pipe axis is greater than $2.5\%$ \added{(see \S \ref{sec:Validity})}. Note that all curves asymptotically approach a vertical line corresponds to $Ri_{s}\simeq 59.86$. In $(b)$, the blue and red curves indicate the trajectories of the two saddle-node points with change in $Q$, respectively. The graph at the bottom shows the projection of the trajectories onto the $Q-Ri$ plane. \added{$(c-e)$ Velocity profile $U(r)$ of the steady solutions along the continuation from $Ri=50$ to the singularity at each $Q$. The colour of the lines changes from green to blue, representing increasing $N(0)$ along the continuation.}}
\end{figure}

\begin{figure}
    \centering
    \includegraphics[width=0.9 \columnwidth]{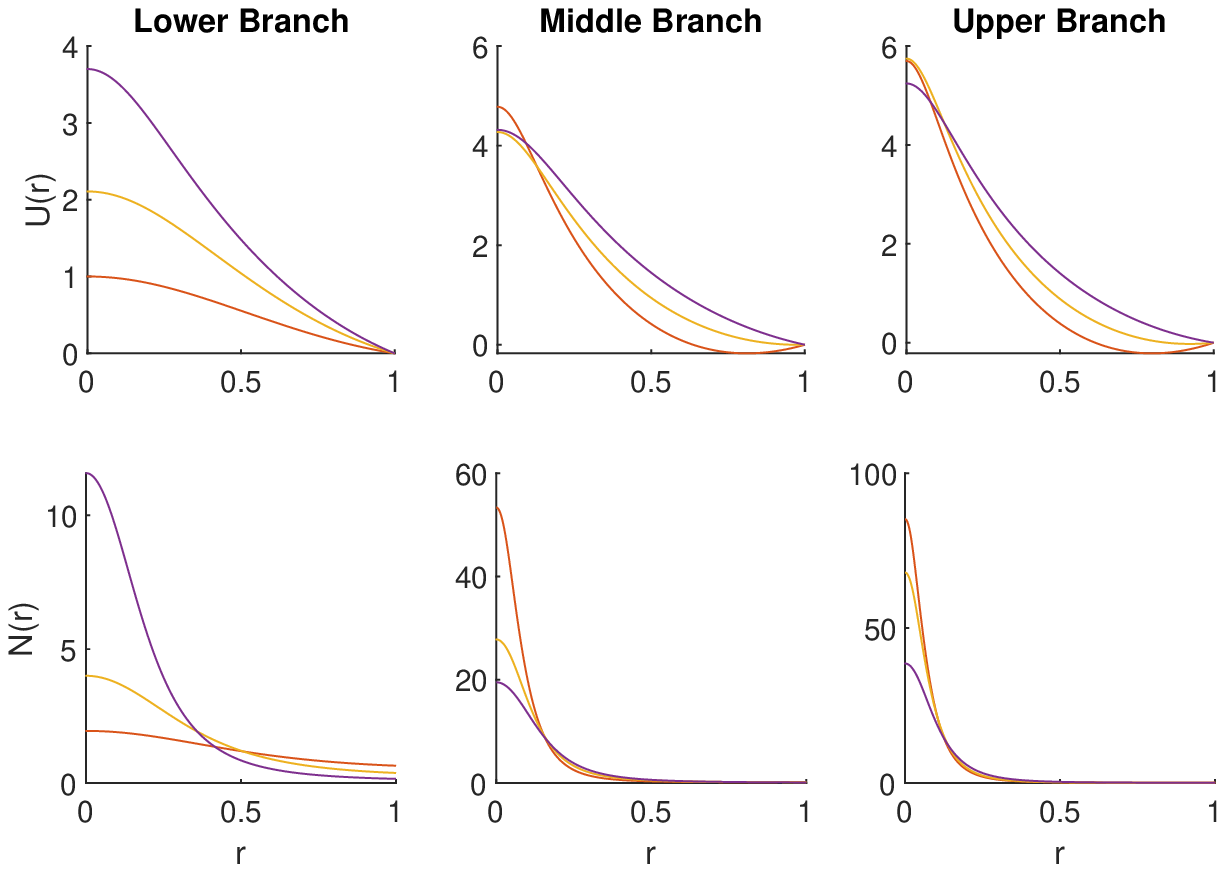}
    \caption{Downward velocity $U(r)$ (top) and cell concentration $N(r)$ (bottom) profiles of the steady basic states at \added{(red) $Q=1.1;Ri=65.9$, (yellow) $Q=2.1;Ri=62$ and (purple) $Q=3.1;Ri=56.9$} (model G). Here, the left, middle and right columns represent the lower-, middle- and upper-branch solutions, respectively. }
    \label{fig:6}
\end{figure}

Now, we compare the steady basic-state solutions obtained from model F to those from model G. Figure \ref{fig:5a} shows the bifurcation diagram of the steady solution obtained using model G. When the cell concentration at the pipe axis is relatively small (i.e. $N(0)<10^2$), the same kind of cusp bifurcation is seen in this case (compare with figure \ref{fig:3a}). Also, the related form of the steady solutions is qualitatively identical to that obtained with model F, as shown in figure \ref{fig:6}. However, as the solution is continued further from the middle branches, its behaviour turns out to be very different from that obtained with model F. In particular, for all the flow rates considered, the bifurcation curves do not properly form the upper branches, contrary to the model F (compare with figure \ref{fig:3b}). Indeed, irrespective of the flow rate $Q$, the continued solution from the middle branch asymptotically exhibits a singular behaviour at $Ri=Ri_{s}(\simeq 59.86)$ in figure \ref{fig:5a}), implying $N(0) \rightarrow \infty$ with the continuation. This suggests that a steady downflowing upper-branch solutions may not necessarily exist for $Ri>Ri_{s}$. 

\begin{figure}
\centering{}
	\sidesubfloat[]{\includegraphics[width = 0.45 \columnwidth]{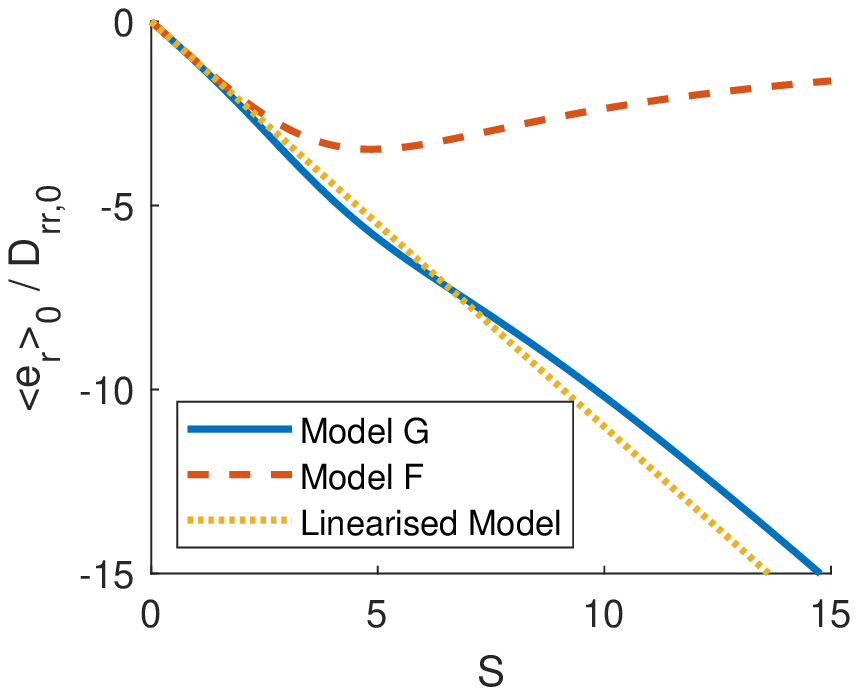}\label{fig:7a}}
	\sidesubfloat[]{\includegraphics[width = 0.45 \columnwidth]{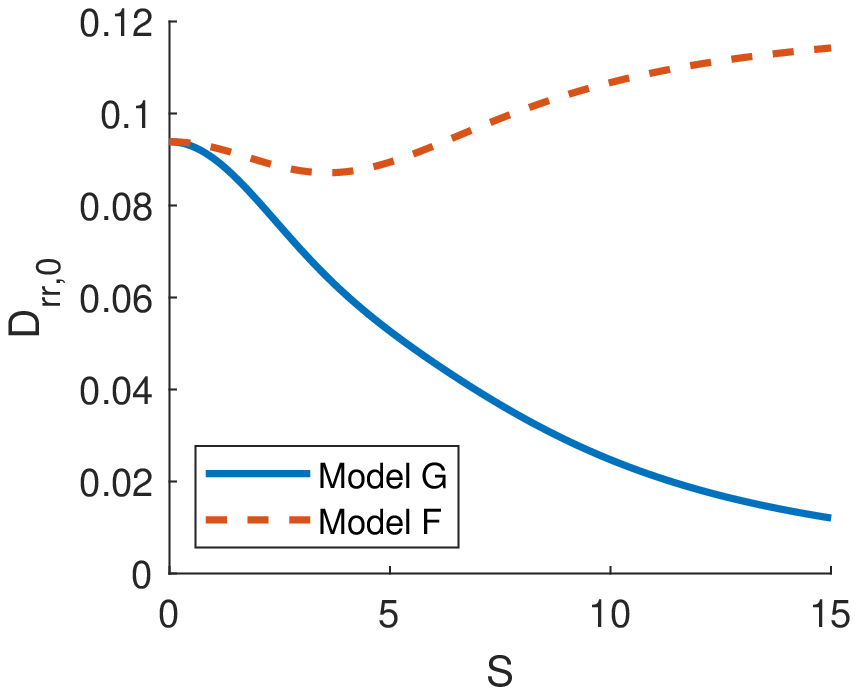}\label{fig:7b}}
\caption{Averaged radial swimming velocity and radial diagonal component of diffusivity tensor with the dimensionless shear rate,  $S=-{D_R}^{-1}{dU}/{dr}$ \cite[see also figure 1][]{Bearon2012}: $(a)$ $\esth{r}_0/D_{rr,0}$; $(b)$ $D_{rr,0}$. Here, \protect \dashed, model F; \protect \solid, model G. In $(a)$, \protect \dotted, linearised model (see \ref{eq:Drr}).} \label{fig:7}
\end{figure}

\begin{figure}
    \vspace*{0.3cm}
	\centering{}
	\sidesubfloat[]{\includegraphics[width = 0.38 \columnwidth]{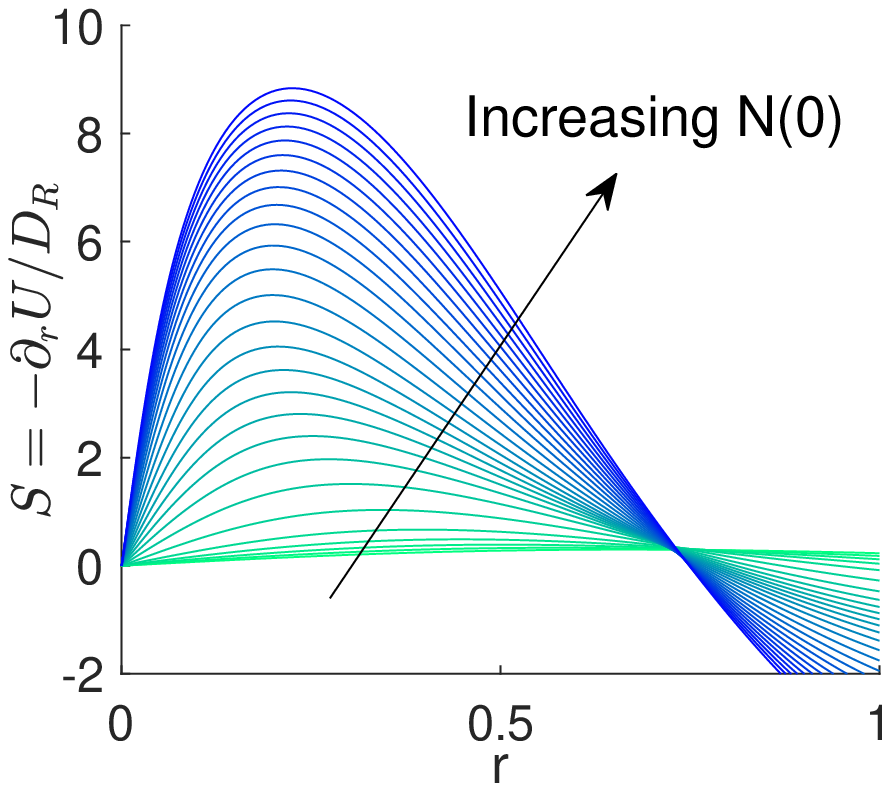}\label{fig:8a}}
	\sidesubfloat[]{\includegraphics[width = 0.38 \columnwidth]{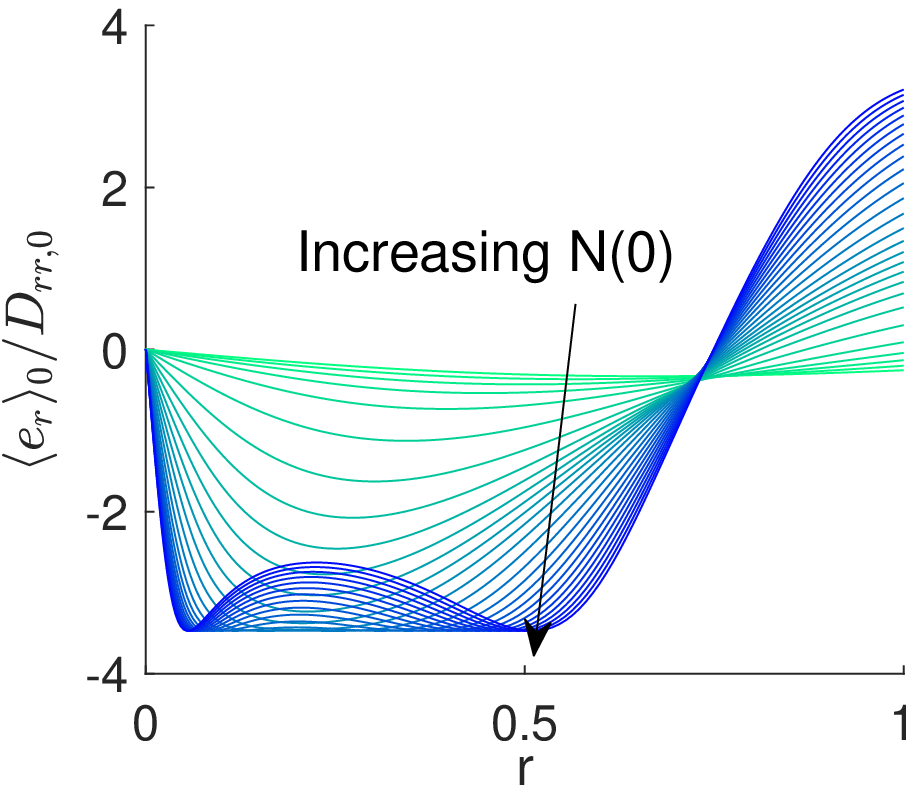}\label{fig:8b}}\\\vspace*{0.2cm}
	\sidesubfloat[]{\includegraphics[width = 0.38 \columnwidth]{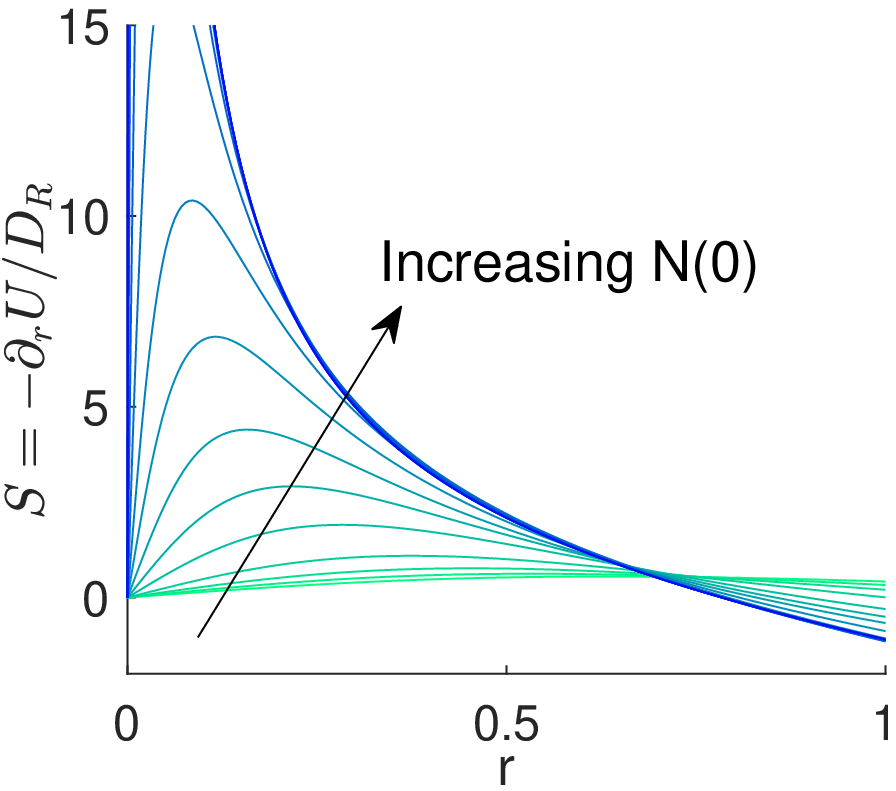}\label{fig:8c}}
	\sidesubfloat[]{\includegraphics[width = 0.38\columnwidth]{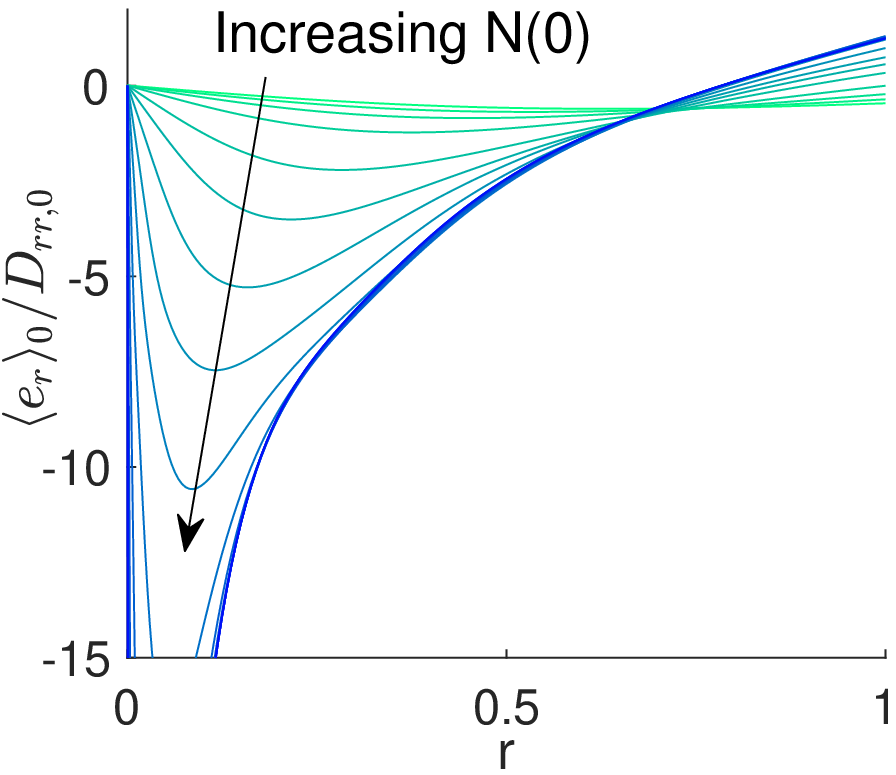}\label{fig:8d}} 
	\caption{\added{($a$,$c$) Rotary P\'{e}clet number $S(r)$ and ($b$,$d$) net-flux-diffusion ratio $\esth{r}_0/D_{rr,0}$ against radial position $r$ along the continuation at $Q=2.1$ in ($a$-$b$) model F and ($c$-$d$) model G.}} \label{fig:8}
\end{figure}

The origin of the singularity in the steady solution can be further studied from the following explicit form of $N(r)$ obtained with (\ref{eq:n0}) and (\ref{eq:vel0_n0_BC}) \cite[][]{Bees2010a}:
\begin{equation}\label{eq:n_base_explicit}
    N(r)= N(0) \exp \Big(D_R \int_0^r \frac{\esth{r}_0}{D_{rr,0}} dr\Big).
\end{equation}
This form of the solution indicates that the singularity would be directly related to the behaviour of $\esth{r}_0/D_{rr,0}$ with prescribed background shear $S(\equiv-1/D_R dU/dr)$. \added{Note that $S$ is varying with the radial position $r$ (figure \ref{fig:8}) because of the coupling via the negative buoyancy. In effect, $S(r)$ can also be interpreted as the local rotary P\'{e}clet number. }Figure \ref{fig:7a} shows the behaviour of $\esth{r}_0/D_{rr,0}$ with the background shear $S$ for both model F and G (see also figure 1$(f)$ in \cite{Bearon2012}). For model F, the values of $\esth{r}_0/D_{rr,0}$ are bounded between a relatively small negative value ($\min(\esth{r}_0/D_{rr,0})=-3.47$) and zero for all the values of $S$. However, in the case of model G, $\esth{r}_0/D_{rr,0}$ monotonically decreases with $S$ and is roughly linearly proportional to $S$ for sufficiently large $S$ \added{(figure \ref{fig:7a})}. It should be mentioned that this difference in $\esth{r}_0/D_{rr,0}$ between model F and G must originate from the difference in the translational diffusivity, because they share exactly the same $\esth{r}$ (see also \S\ref{sec:Problem_Formulation}). This is shown in figure \ref{fig:7b}.

\added{Figure 8 visualises the radial profiles of $S(r)$ and $\esth{r}_0/D_{rr,0}(r)$ for model F and G along the continuation curves at $Q=2.1$. As the solution is continued from the lower to the upper branch, the absolute values of $S(r)$ monotonically increase for both model F and G (figures \ref{fig:8}$a$, $b$) However, $\esth{r}_0/D_{rr,0}(r)$ do not behave like $S(r)$. In particular, $\esth{r}_0/D_{rr,0}(r)$ for model F is bounded due to the nature shown in figure \ref{fig:7a}. As a result, $\esth{r}_0/D_{rr,0}(r)$  in figure \ref{fig:8b} is bounded, whereas that in \ref{fig:8d} is not. This suggests that the singularity developed at $Ri_s$ from model G originates from the unboundedness of $\esth{r}_0/D_{rr,0}(S)$.} 

\begin{figure}
    \centering
    \includegraphics[width = 0.75 \columnwidth]{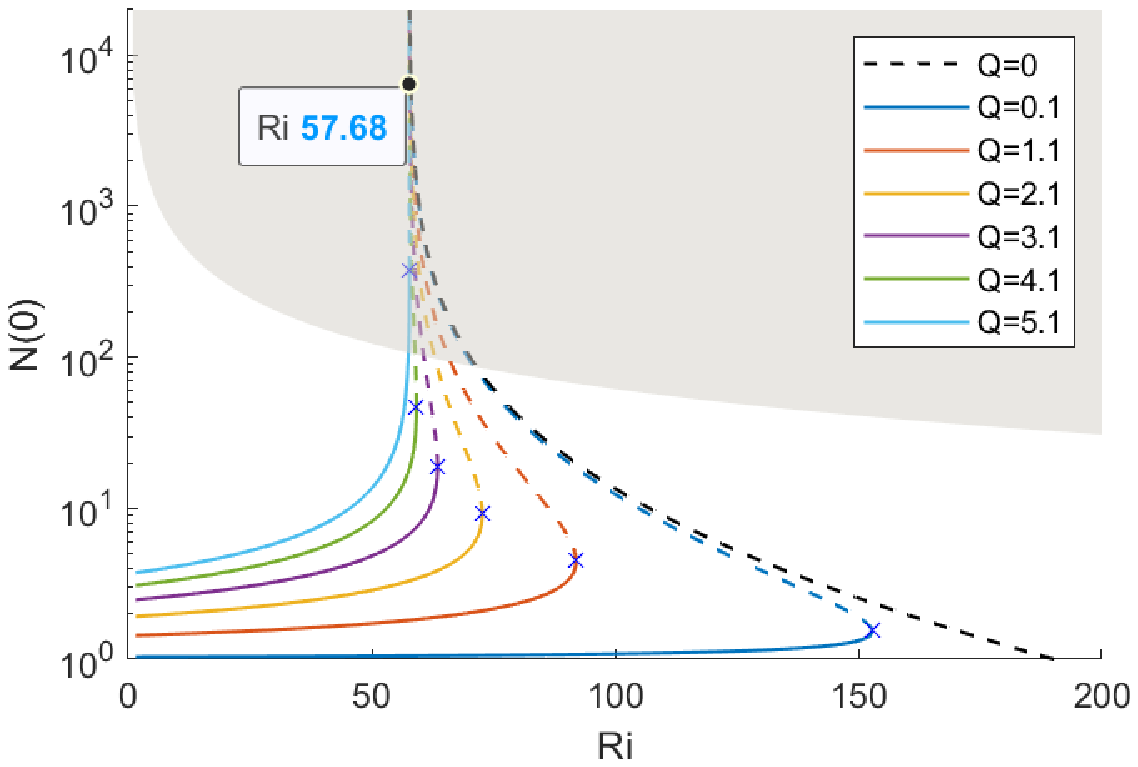}
    \caption{Bifurcation diagram of the linearised model with $Ri$ for several $Q$. Here, the state of steady solution is represented by the cell concentration at the pipe axis $N(0)$. As the solution is continued from the lower branch, $N(0) \rightarrow \infty$ at $Ri=Ri_s(\simeq 57.68)$. Also, \protect \solid, stable; \protect \dashed, unstable, and the blue crosses (x) indicate the saddle-node point. The grey area indicates the cases where the volume fraction at the pipe axis is greater than $2.5\%$ \added{(see \S \ref{sec:Validity}).} }
    \label{fig:9}
\end{figure}{}

If the behaviour of $\esth{r}_0/D_{rr,0}$ in model G is further simplified, the $Ri_s$ in figure \ref{fig:5b} can be predicted analytically. For this purpose, we consider a simplified model for $\esth{r}_0/D_{rr,0}$, in which the translational diffusivity is set such that $\esth{r}_0/D_{rr,0}$ is linearly proportional to the background shear $dU/dr$:
\begin{equation} \label{eq:Drr}
    D_{rr,0}= -\frac{\esth{r}_0 D_R}{\eta}\left (\frac{dU}{dr} \right)^{-1}\added{=\frac{\esth{r}_0}{\eta S}}. 
\end{equation}
In this model, $\esth{r}_0$ is kept to be the same as that used in both model F and G, and $\eta$ is obtained from the slope of $\esth{r}_0/D_{rr,0}$ for $S=0$: i.e. $\eta\equiv-\partial (\esth{r}_0/D_{rr,0})/\partial S|_{S=0}$. The behaviour of this simplified model is shown in figure \ref{fig:7a}, in which $\esth{r}_0/D_{rr,0}$ of this model behaves similarly to that of model G with $\eta=1.10$. We shall refer to this model as the `linearised model'. 

Coincidentally, the behaviour of $\esth{r}_0/D_{rr,0}$ against $S$ in this `linearised model' is the same as that of \cite{Kessler1986}. In \cite{Kessler1986}, $\esth{r}_0$ is linear to shear rate $S$ and $D_{rr,0}$ is constant. In contrast, in both model G and the subsequent `linearised model', $D_{rr,0} \sim S^{-2}$ and $\esth{r}_0 \sim S^{-1}$ as $S$ increases towards infinity.  As a result, the `linearised model' shares the same linear behaviour in $\esth{r}_0/D_{rr,0}(S)$ with the simpler model of \cite{Kessler1986}, even though their approximations for $\esth{r}_0$ and $D_{rr,0}$ respectively are different. In the following, we will exploit the similiarity in $\esth{r}_0/D_{rr,0}(S)$ between the `linearised model' and \cite{Kessler1986} to understand the bifurcation.

Bifurcation of the steady solutions of the linearised model is shown in figure \ref{fig:9} for $\eta=1.10$. The bifurcation diagram of the linearised model confirms the behaviour qualitatively similar to that of model G: as the solution is continued from the lower to upper branch, the cell concentration at the pipe axis gradually becomes singular at $Ri\simeq Ri_s(=57.68)$ irrespective of $Q$ (compare with figure \ref{fig:5a}). The only qualitative difference between figure \ref{fig:5a} and figure \ref{fig:9} is the small extra bumpy behaviour in the bifurcation curves of model G, likely the consequence of the nonlinear behaviour of $\esth{r}_0/D_{rr,0}$ in this model. This suggests that the singularity at the centreline cell concentration in model G indeed originates from the monotonically decreasing $\esth{r}_0/D_{rr,0}$ with the background shear rate $S$. In fact, as $S \rightarrow \infty$, $\esth{r}_0$ scales with $S^{-1}$ while $D_{rr,0}$ scales with $S^{-2}$, hence $\esth{r}_0/D_{rr,0}$ scales with $S$ \citep{Bearon2012}. This implies that the origin of the singularity in $N(0)$ is essentially associated with the lack of translational diffusion flux relative to the advection flux caused by swimming in the radial direction in model G, as the local shear rate $S$ increases with $N(0)$ towards infinity along the continuation.

The emergence of this singularity can be more precisely analysed. We now consider the following equation for $N(r)$, which can be obtained from (\ref{eq:base_full}) with $D_{rr,0}$ in (\ref{eq:Drr}) of the linearised model:
\begin{equation}\label{eq:N_G}
    -\frac{1}{r}\frac{d}{dr}\Big(r\frac{d}{dr} \ln{N(r)}\Big)=8 \gamma \added{(N(r)-1) -} G,
\end{equation} 
where $\gamma={\eta RiRe}/8$ and \replaced{$G= \eta \partial U/\partial r|_{r=1}$}{$G=-\Rey \eta \partial P_0/\partial z$}. Since $N(r)$ near the singular regime is highly concentrated near the pipe axis, we assume that the cell concentration is highly focused in a small region around the axis: i.e. $r\in[0,\epsilon]$. Then, from the constraint of $N(r)$ given by (\ref{eq:n_norm}), $N(r) \sim O(\epsilon^{-2})$ in $r\in[0,\epsilon]$. Therefore, in this region, we can introduce a relevant rescaling of the radial coordinate $r=\epsilon R$, where $R$ is of order unity, and define a normalised profile $N_0(R)=N(r/\epsilon)/N(0)$. Then, at $O(\epsilon^{-2})$, (\ref{eq:N_G}) is approximated as
\begin{equation}\label{eq:N_G1}
\left(\frac{1}{N_0}\frac{dN_0}{dR}\right)^2-\frac{1}{RN_0}\frac{dN_0}{dR}-\frac{1}{N_0}\frac{d^2N_0}{dR^2}={8\gamma} N_0(R).
\end{equation}
Now, it is evident that (\ref{eq:N_G1}) does not contain $G$ anymore, although it still describes the behaviour of most of the cell concentration in the suspension. This implies that, in the regime where the singular solution nearly develops, the form of the steady basics-state solution should approximately be independent of pressure gradient as well as of flow rate for  $r\in[0,\epsilon]$, confirming the numerical result for $N(0)>O(10^3)$ shown in figure \ref{fig:9}. 

\begin{figure}
    \centering
    \sidesubfloat[]{
    \includegraphics[width = 0.95 \columnwidth]{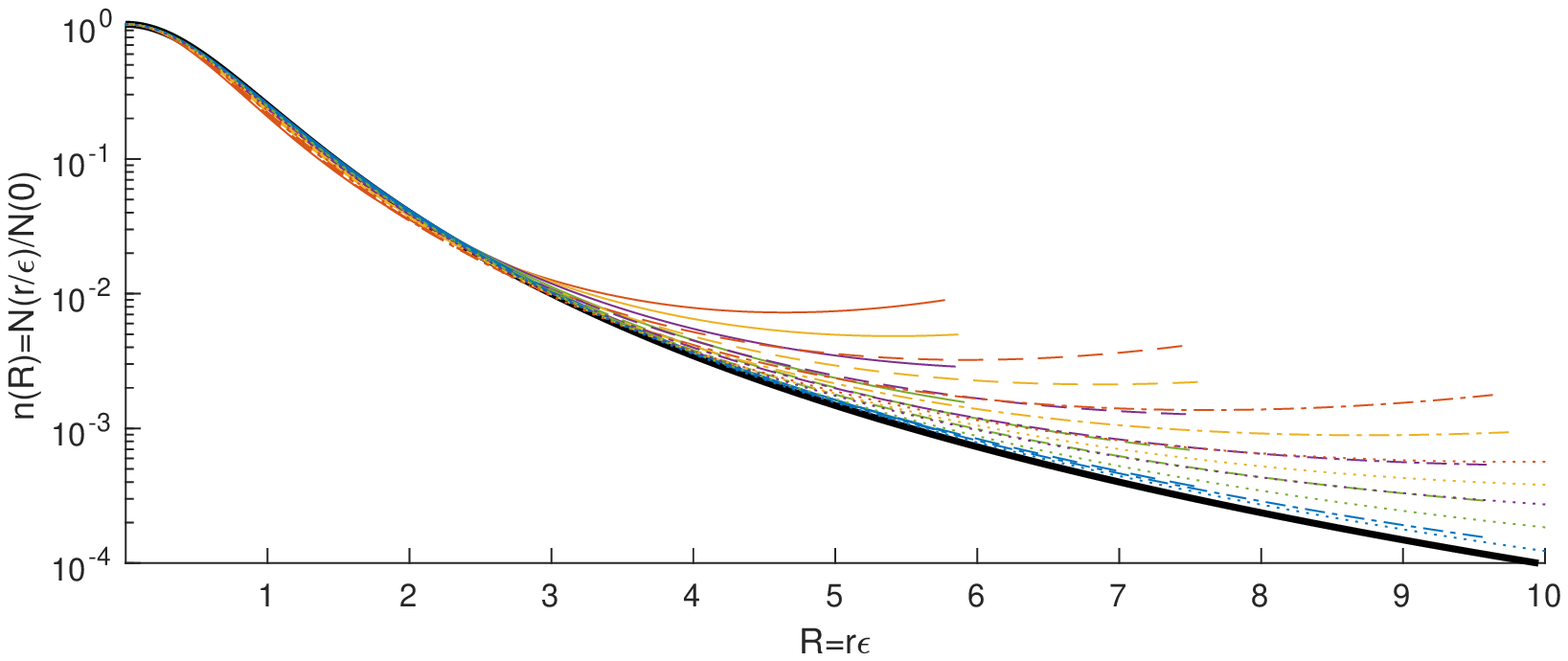}}\\
    \sidesubfloat[]{
    \includegraphics[width = 0.95 \columnwidth]{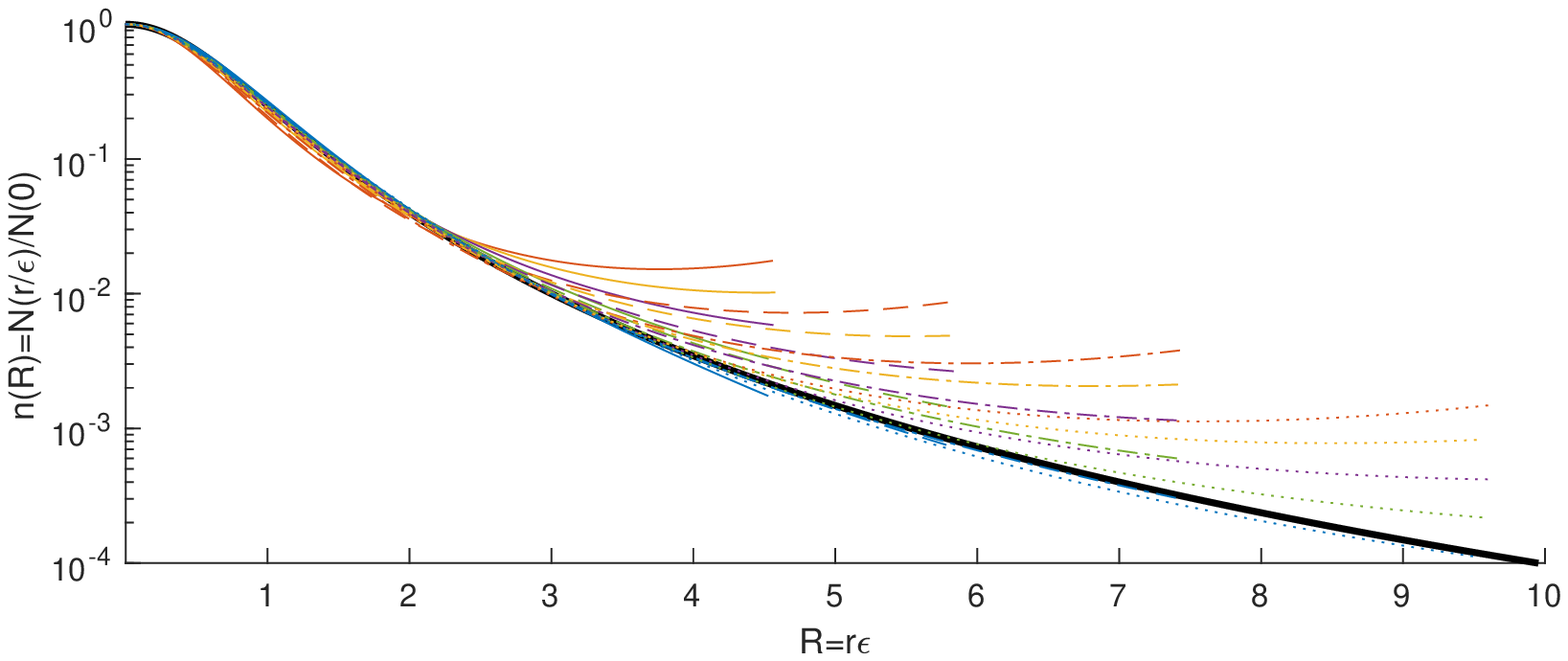}}
    \caption{The cell concentration profile of the steady solutions noramlised by the pipe axis value for several flow rates ($Q=1,2,3,4,5$): $(a)$ the linearised model; $(b)$ model G. Here, $N(0)=20$, \protect\tikz[baseline]{\protect\draw[line width=0.5mm] (0,.8ex)--++(1,0) ;};
    $N(0)=33$, \protect\tikz[baseline]{\protect\draw[line width=0.5mm,dashed] (0,.8ex)--++(1,0) ;};
    $N(0)=55$, \protect\tikz[baseline]{\protect\draw[line width=0.5mm,dash dot] (0,.8ex)--++(1,0) ;};
    $N(0)=90$, \protect\tikz[baseline]{\protect\draw[line width=0.5mm,dotted] (0,.8ex)--++(1,0) ;} for the coloured lines. The thick black solid line indicates the solution of \cite{Kessler1986} given in  (\ref{eq:Kessler_analytic}).}
    \label{fig:10}
\end{figure}

The solution to (\ref{eq:N_G1}) was previously obtained by \cite{Kessler1986} with the far field boundary condition $N_0(\infty)=0$: i.e.
\begin{equation}\label{eq:Kessler_analytic}
    N_0(R)=\frac{1}{(1+\gamma R^2)^2}. 
\end{equation} 
We note that (\ref{eq:N_G1}) is the leading-order approximation of (\ref{eq:N_G}) valid only for $R\in[0,1]$ (or equivalently $r\in[0,\epsilon]$). Therefore, (\ref{eq:Kessler_analytic}) would be a good approximation of the full numerical solution of (\ref{eq:N_G}) for any $Q$ in this region. The normalised cell concentration of its numerical solutions for several $Q$ and $N(0)$ is compared with (\ref{eq:Kessler_analytic}) in figure \ref{fig:10}$(a)$. Indeed, (\ref{eq:Kessler_analytic}) shows an excellent agreement with the numerical solutions for $R\in[0,1]$. We note that (\ref{eq:Kessler_analytic}) is also a good approximation of the cell-concentration profile of model G near the pipe axis, as demonstrated in figure \ref{fig:10}$(b)$.

Finally, if the solution (\ref{eq:Kessler_analytic}) is substituted into (\ref{eq:n_norm}), the resulting cell concentration at the pipe axis is obtained as
\begin{equation}
N(0)=\frac{1}{1-\gamma}. 
\end{equation}
From the definition of $\gamma$, this implies that $N(0)$ would be singular, if $\gamma\rightarrow 1$. At $\gamma=1$, $Ri=8/(\eta Re)$, which should be a good approximation of $Ri_s$. Indeed, for the given $Re$ and $\eta$ of the linearised model, the value of $8/(\eta Re))$ is $57.66$, showing excellent agreement with the numerical one $Ri_s\simeq 57.68$ (see figure \ref{fig:9}). 

\added{The physical mechanism for the singularity is as follow. As $Ri$ increases, the acceleration from the negative buoyancy of cells would increase local downflow and the shear rate $S$, attracting more cells towards the centre from more negative $\langle e_r \rangle_0/D_{rr,0}$ for the increased $S$. The increased cell concentration would therefore increase the local downflow further, and the same process would be repeated until the resulting averaged swimming flux and the radial diffusion are balanced (i.e. formation of a steady solution). Therefore, $N(0)$ would increase drastically with a small increment in $Ri$. The physical process described here is identical to that of the gyrotactic instability. However, in the current case as well as that of \cite{Kessler1986}, there is a certain threshold of $Ri(=Ri_s)$, beyond which the formation of steady solution is no longer possible. }

\begin{figure}[ht]
\centering{}
	\sidesubfloat[]{\includegraphics[width = 0.4 \columnwidth]{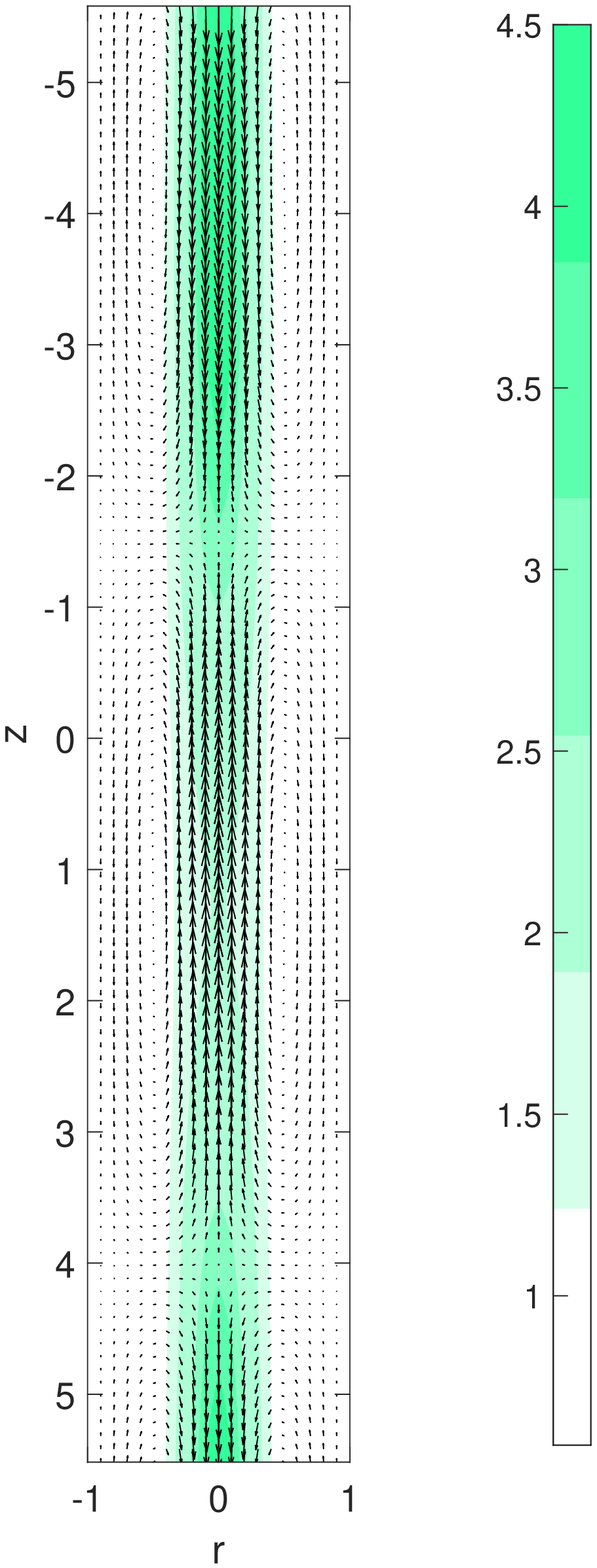}}
	\sidesubfloat[]{\includegraphics[width = 0.4 \columnwidth]{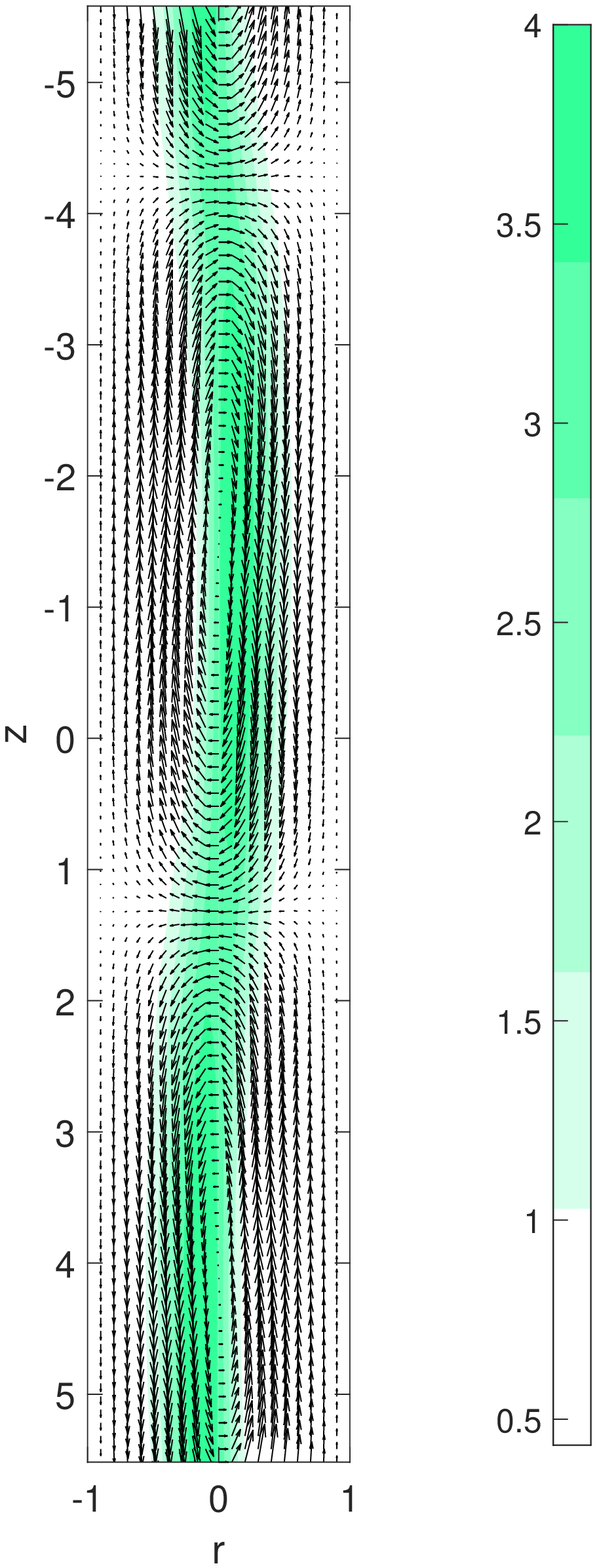}}
\caption{Spatial structure of the unstable $(a)$ axisymmetric ($m=0$) and $(b)$ non-axisymmetric ($m=1$) mode at $Ri=92, Q=1, \alpha=0.56$, computed using a lower-branch basic state. Here, the contours indicate $N(r)+a n'(r,z)$ with an arbitrary value of $a$ for visualisation, and the vectors represent the axial and radial perturbation velocity field.\label{fig:11}}
\end{figure}

\section{Linear stability\label{sec:Results_stability}}
Now, linear stability analysis is performed with the steady plume solutions computed in \S \ref{sec:Results_basic_state}. Here, we will focus on the axisymmetric mode and the first non-axisymmetric mode, which are similar to the varicose and sinuous modes in downward channel flow \citep{HP2014b}. The typical spatial structures of these two modes are visualised in figure \ref{fig:11}, demonstrating the similarity to figure 8 in \cite{HP2014b}: the axisymmetric mode is composed of a plume, the thickness of which varies along the axial direction, whereas the non-axisymmetric mode is composed of an anaxially meandering plume. In the present study, we will be focusing on the axisymmetric mode ($m=0$) as it is more physically relevant and observable in experiments. The first non-axisymmetric mode (i.e. when $m=1$) is also presented briefly. 

Given the existence of multiple basic state solutions for a given set of $Ri$ and $Q$, it is not straightforward to present the conventional neutral stability diagram. Therefore, in this section, we present the stability of the steady solutions along each of the continuation curves shown in \S\ref{sec:Results_basic_state}. For each steady solution (basic state) and the corresponding set of the parameters, the maximum growth rate $\omega_{i,max}$ is sought out for all real $\alpha$. Here, the value of $\alpha$ used to search for $\omega_{i,max}$ ranges from $0.001$ to $20$, and the corresponding $\alpha_{max}$ is also computed. 

\subsection{Axisymmetric mode}
\begin{figure}
\centering{}
	\sidesubfloat[]{\includegraphics[width = 0.45 \columnwidth]{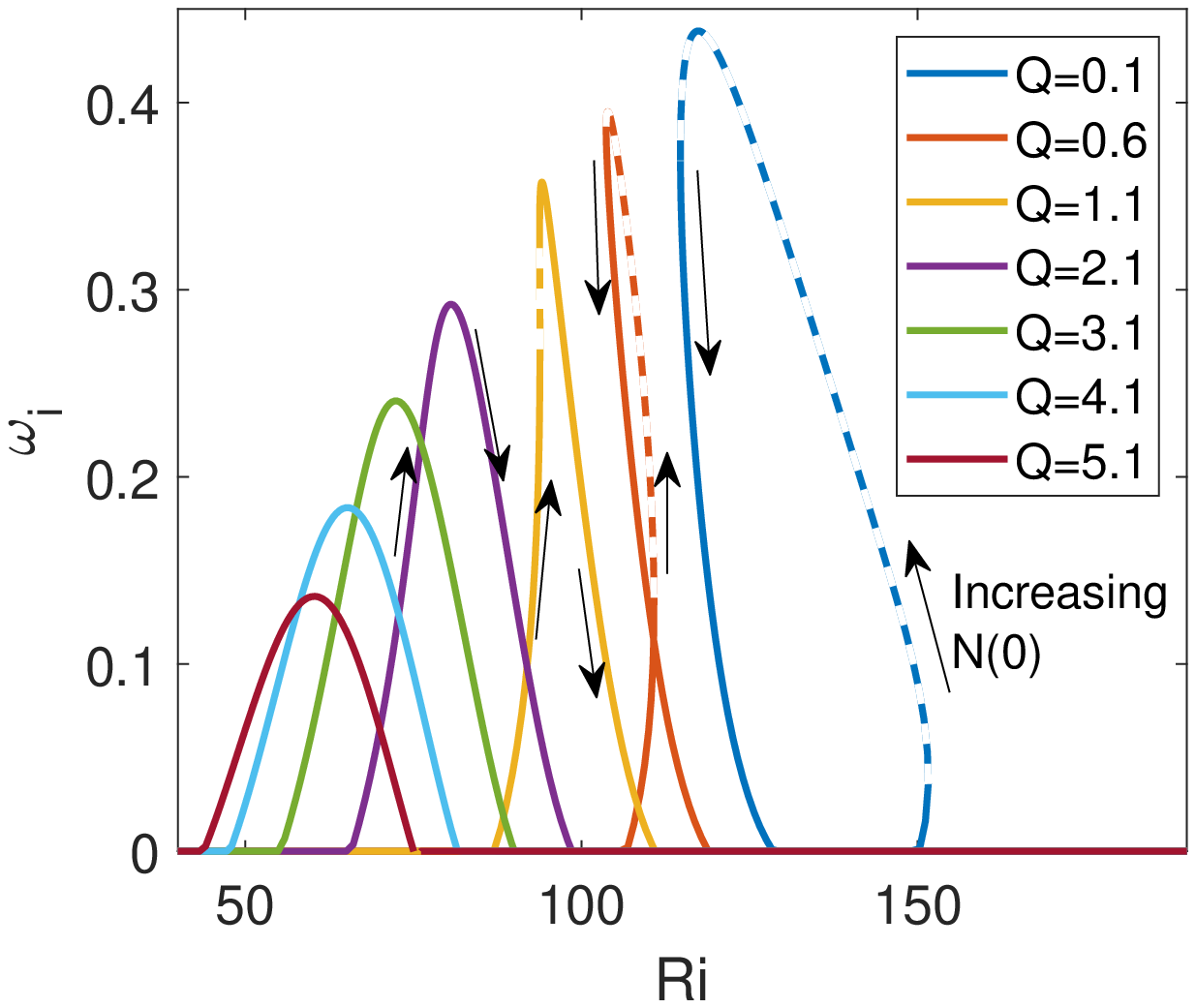}\label{fig:12a}}
	\sidesubfloat[]{\includegraphics[width = 0.45 \columnwidth]{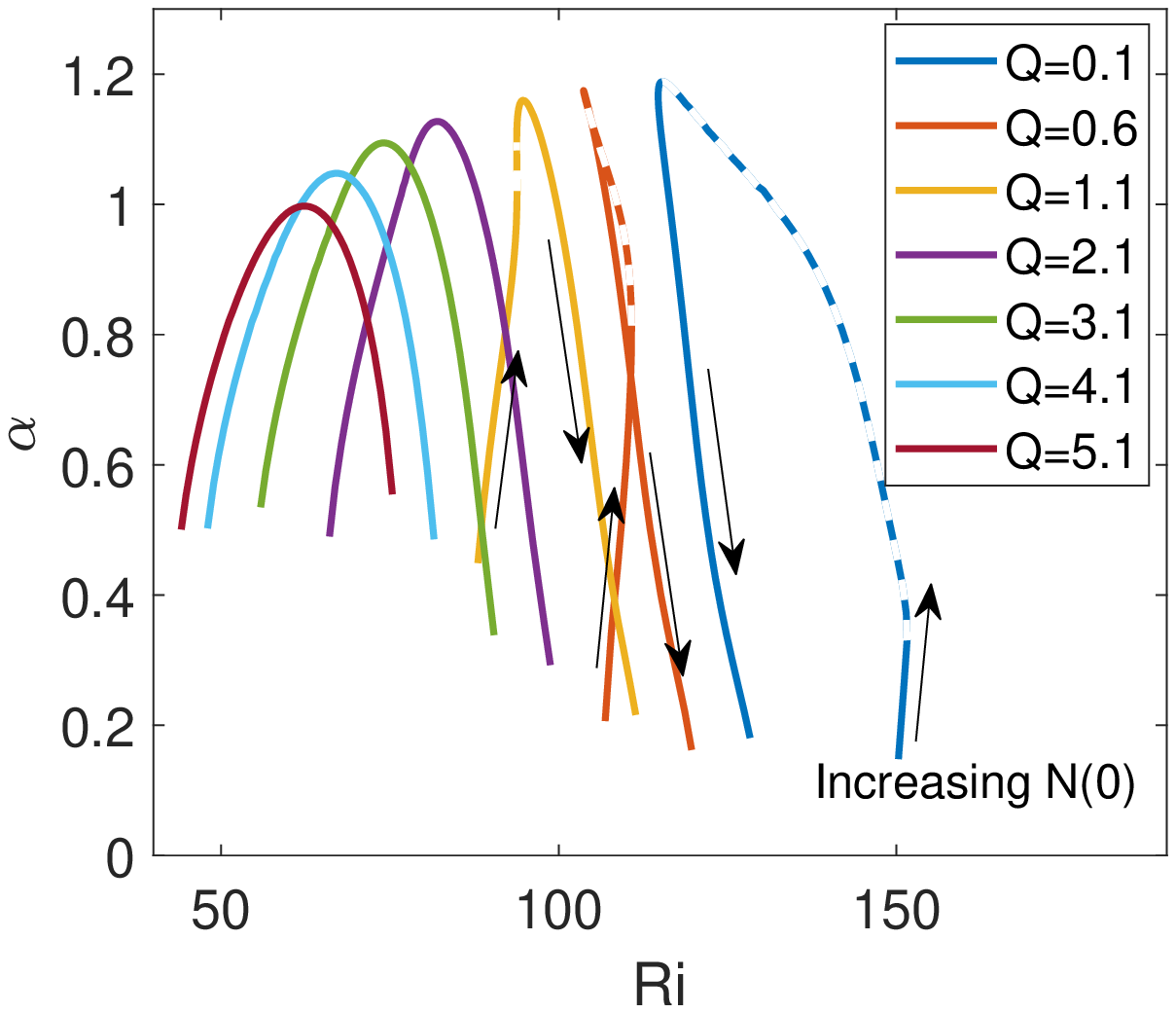}\label{fig:12b}}
\caption{Stability of the axisymmetric mode (model F): $(a)$ maximum growth rate $\omega_{i,max}$ and $(b)$ the corresponding $\alpha_{max}$ of the steady solution at each $Ri$ and $Q$. In $(a)$, \protect \solid, stable; \protect \dashed, unstable to streamwise-uniform perturbation.}\label{fig:12}
\end{figure}
\subsubsection{Model F}
Figure \ref{fig:12} shows how the maximum growth rate $\omega_{i,max}$ and the corresponding streamwise wavenumber $\alpha_{max}$ change with $Ri$ for each fixed $Q$ along the continuation curves in figure \ref{fig:3a}. At low $Q(\lesssim 1)$, the basic state becomes unstable to axially varying perturbations near the first saddle-node point where $N(0)$ increases rapidly with $Ri$, as the solution continued from the lower to the middle branch. With a further continuation, the solution stabilises, in that the value of $\omega_{i}$ begins to reduce, before the second saddle-node point (figure \ref{fig:12a}). The streamwise wavenumber retaining the maximum growth rate also behaves similarly to  $\omega_{i,max}$: $\alpha_{max}$ grows as the solution continued from the lower to the middle branch, and it decays again with a further continuation. Finally, as $Q$ increases, $\omega_{i,max}$ obtained for all the steady solutions along the continuation decreases, implying that increasing the flow rate stabilises the streamwise perturbation. 

Comparison of figure \ref{fig:12a} with figure \ref{fig:3a} also indicates that the destabilisation seems to correlate with the rapid increase in $N(0)$ of the basic state at least for the lower and middle branches. This suggests that the instability is presumably associated with the sharp gradient in the base flow and cell concentration near the pipe axis, consistent with the previous observation in \cite{HP2014b} where the instability of this type \cite[varicose mode][]{HP2014b} was shown to originate from the following simplified process:
\begin{equation}\label{eq:instm}
    \frac{\partial n'}{\partial t} \sim - n'\Big(\frac{\partial \langle e_r \rangle_0}{\partial r}+\frac{\esth{r}_0}{r}\Big).
\end{equation}
This process also appears through the first to third terms in (\ref{eq:n_pert}) in the present pipe flow. We note that $\partial \langle e_r \rangle_0/\partial r$ and ${\esth{r}_0}/{r}$ must be negative near the pipe axis, because the cells swim towards the centre ($\langle e_r \rangle_0<0$) near the pipe axis and $\langle e_r \rangle_0=0$ at the pipe axis, implying that the same mechanism of the instability is active in the vertical pipe case.  

\begin{figure}
\centering{}
	\sidesubfloat[]{\includegraphics[width = 0.45 \columnwidth]{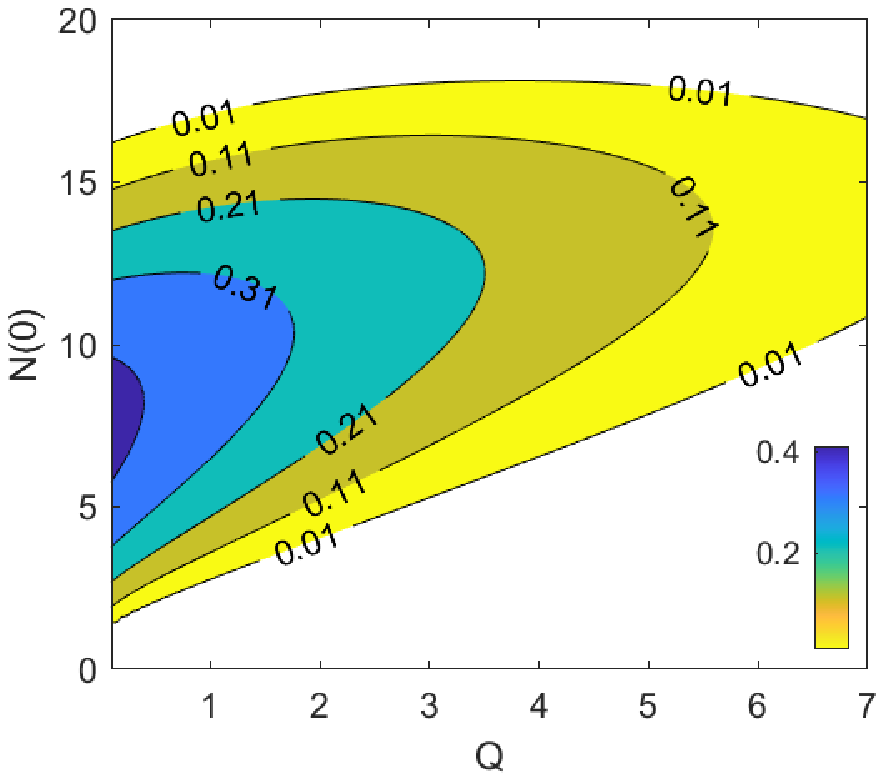}}
	\sidesubfloat[]{\includegraphics[width = 0.45 \columnwidth]{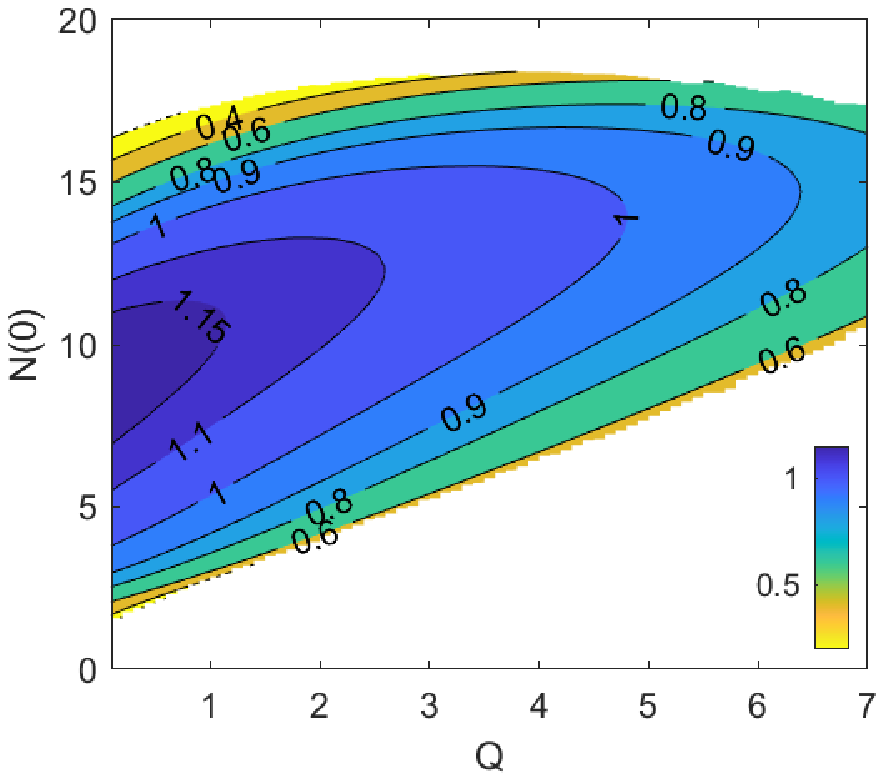}}
\caption{$(a)$ The maximum growth rate $\omega_{i,max}$ and $(b)$ the corresponding $\alpha_{max}$ in the $N(0)-Q$ plane (model F). Note that the contours do not contain any overlap because the basic state is stable before $N(0)$ decreases in the branch continuation.}\label{fig:13}
\end{figure}

To further explain the correlation between the growth rate and the nature of the steady solution near the pipe axis, the maximum growth rate $\omega_{i,max}$ in the $Q-N(0)$ plane is plotted in figure \ref{fig:13}. For $N(0)\lesssim 10$, the basic state is destabilised on increasing $N(0)$, consistent with the explanation given above. However, as $N(0)$ is further increased ($N(0)\gtrsim 10$), the solution is found to be stabilised again. Here, we note that there are no overlapping contour lines in figure \ref{fig:13}, which one might have expected from the existence of multiple $N(0)$ for a given flow rate (see figure \ref{fig:3a}). This is because the basic state is stabilised ($\omega_i<0$) well before $N(0)$ starts to decrease along the upper branch (see figure \ref{fig:3a}). In other words, the stabilisation takes place while $N(0)$ is still increasing along the continuation. 

The stability analysis result here is qualitatively similar to that in \cite{HP2014b}, although the stability diagram in the $Q-Ri$ space is not directly shown here due to the complexity emerging from the bifurcation of basic state: the most unstable mode of the flow appears in the form of an axisymmetric blip instability like their varicose mode, and this instability is stabilised if the flow rate is sufficiently large. However, it should be mentioned that \cite{HP2014b} did not explore for $Ri>90$. Therefore, it is unclear whether the stabilisation observed at higher $N(0)$ in the present pipe flow would also occur in their channel flow. 

\begin{figure}
\centering{}
	\sidesubfloat[]{\includegraphics[width = 0.45 \columnwidth]{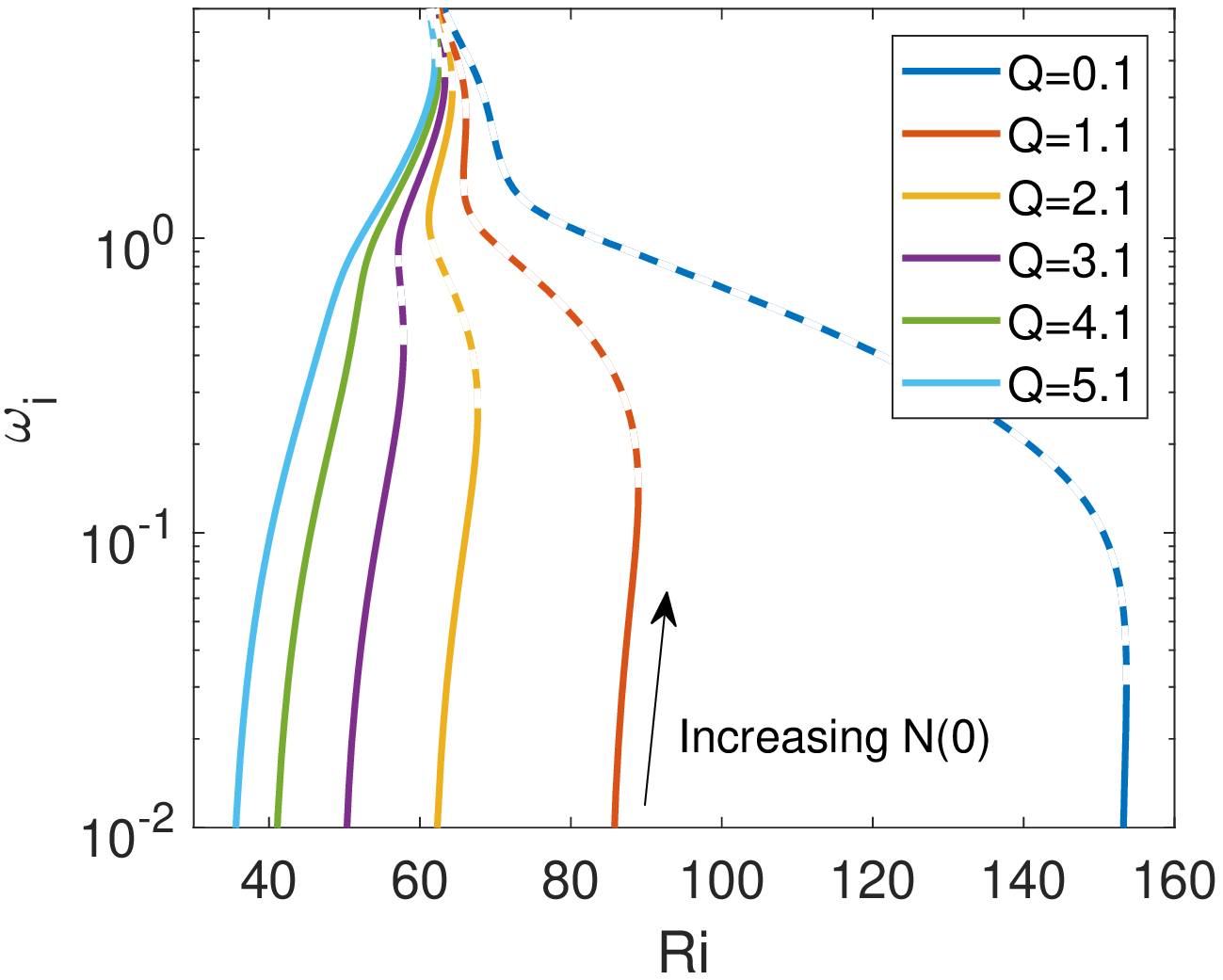}\label{fig:14a}}
	\sidesubfloat[]{\includegraphics[width = 0.45 \columnwidth]{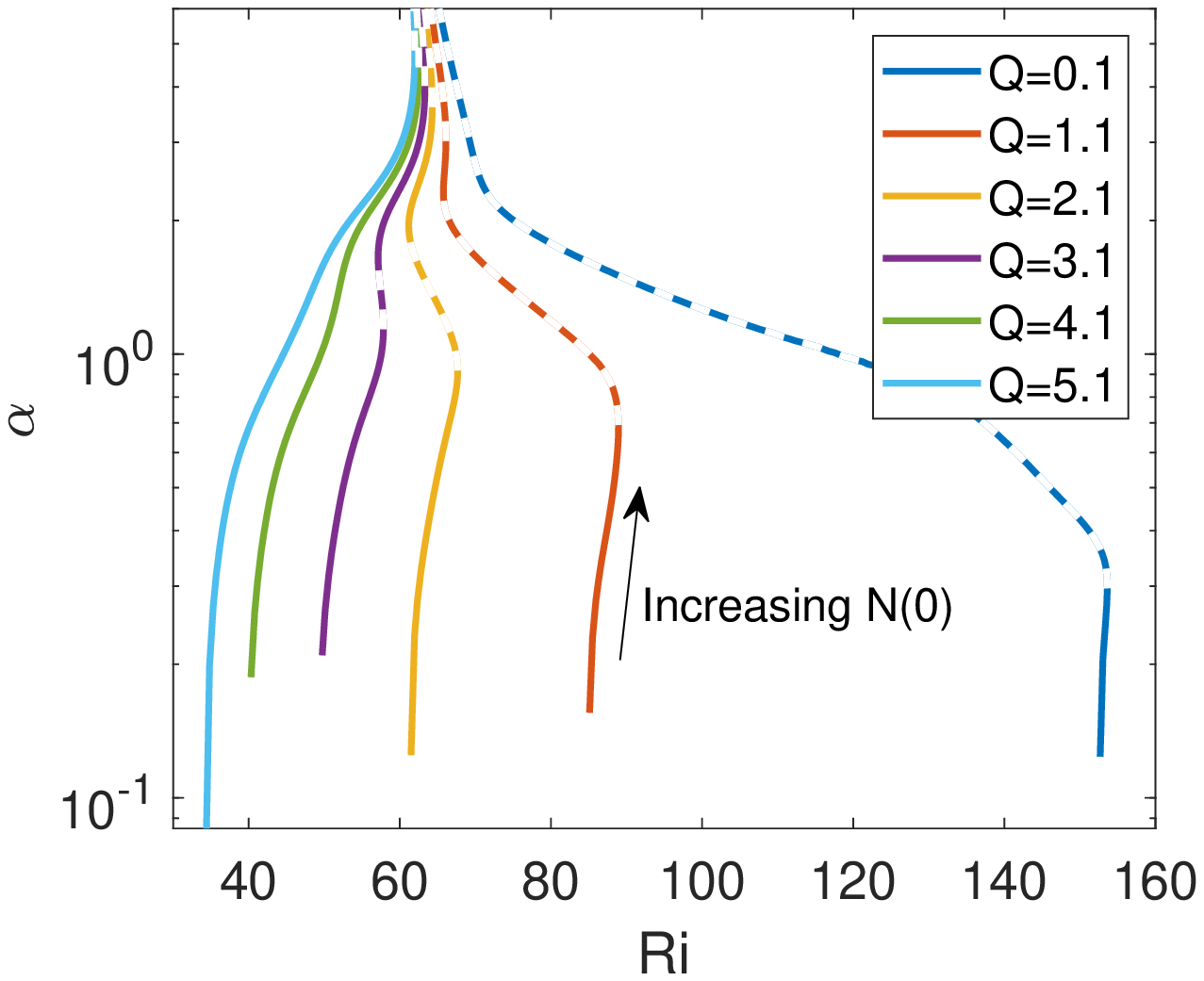}\label{fig:14b}}
\caption{Stability of the axisymmetric mode (model G): $(a)$ maximum growth rate $\omega_{i,max}$ and $(b)$ the corresponding $\alpha_{max}$ of the steady solution at each $Ri$ and $Q$. In $(a)$, \protect \solid, stable; \protect \dashed, unstable to streamwise-uniform perturbation.}\label{fig:14}
\end{figure}

\subsubsection{Model G}
Now, we perform a linear stability using model G. Figure \ref{fig:14} shows how the maximum growth rate $\omega_{i,max}$ changes with $Ri$ along the continuation curves in figure \ref{fig:5a} for each fixed $Q$. Due to the wider range of the values of $\omega_{i,max}$ emerging in model G, here we have chosen to plot $\omega_{i,max}$ in log scale in figure \ref{fig:14}. Similar to the result of model F, the basic state is destabilised near the first saddle-node point. However, in contrast with figure \ref{fig:12}, the basic state is no longer stabilised with the continuation to upper branches. Instead, it is found that the growth rate continues to increase, and the basic state always remains unstable along the continuation curve. 

\begin{figure}
\centering{}
	\sidesubfloat[]{\includegraphics[width = 0.45 \columnwidth]{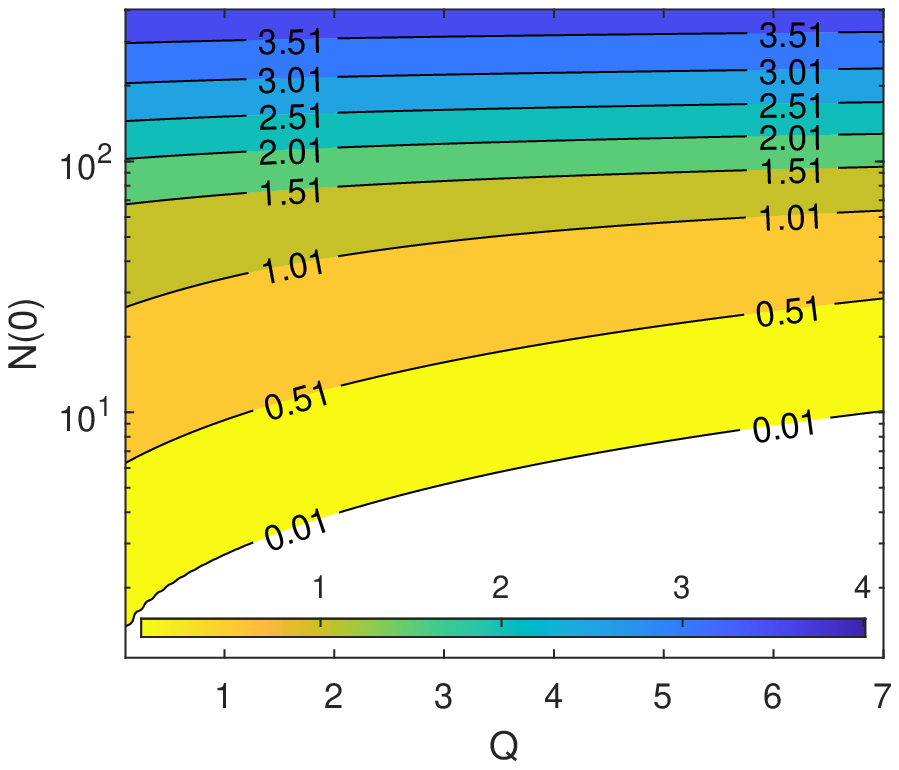}}
	\sidesubfloat[]{\includegraphics[width = 0.45 \columnwidth]{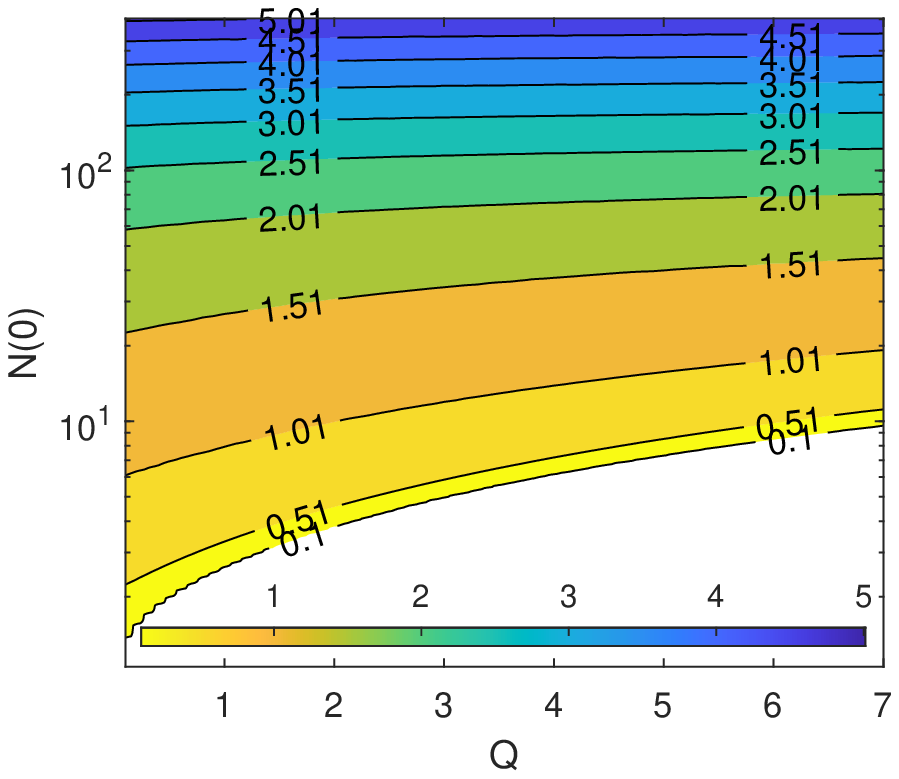}}
\caption{$(a)$ The maximum growth rate $\omega_{i,max}$ and $(b)$ the corresponding $\alpha_{max}$ in the $N(0)-Q$ plane (model G).}\label{fig:15}
\end{figure}

The maximum growth rate $\omega_{i,max}$ in the $Q-N(0)$ space is also plotted in figure \ref{fig:15}. It is interesting to note that the growth rate appears to be almost independent of $Q$ at high enough $N(0)$. This is presumably because the profile of the basic state solution is almost independent of $Q$ for sufficiently high $N(0)$, as discussed in \S \ref{subsec:Bifur_GTD_K}. This observation is also consistent with the notion that the instability is essentially driven by the local flow dynamics near the centreline of the pipe through (\ref{eq:instm}).

The main difference in the stability of model F and G appears from the upper-branch basic state. In a way, this would not be surprising because model F and G show significantly different upper-branch states. Model F exhibits decreasing $N(0)$, as the steady basic-state solution is continued along the upper branch. By the contrary, model G shows increasing $N(0)$ with the continuation. This feature greatly impacts on the stability result, and we shall provide a further discussion on this issue in \S\ref{sec:F_G_Comparison}. 

\subsection{Non-axisymmetric mode \label{sec:sinuous}}
\begin{figure}
\centering{}
	\sidesubfloat[]{\includegraphics[width = 0.45 \columnwidth]{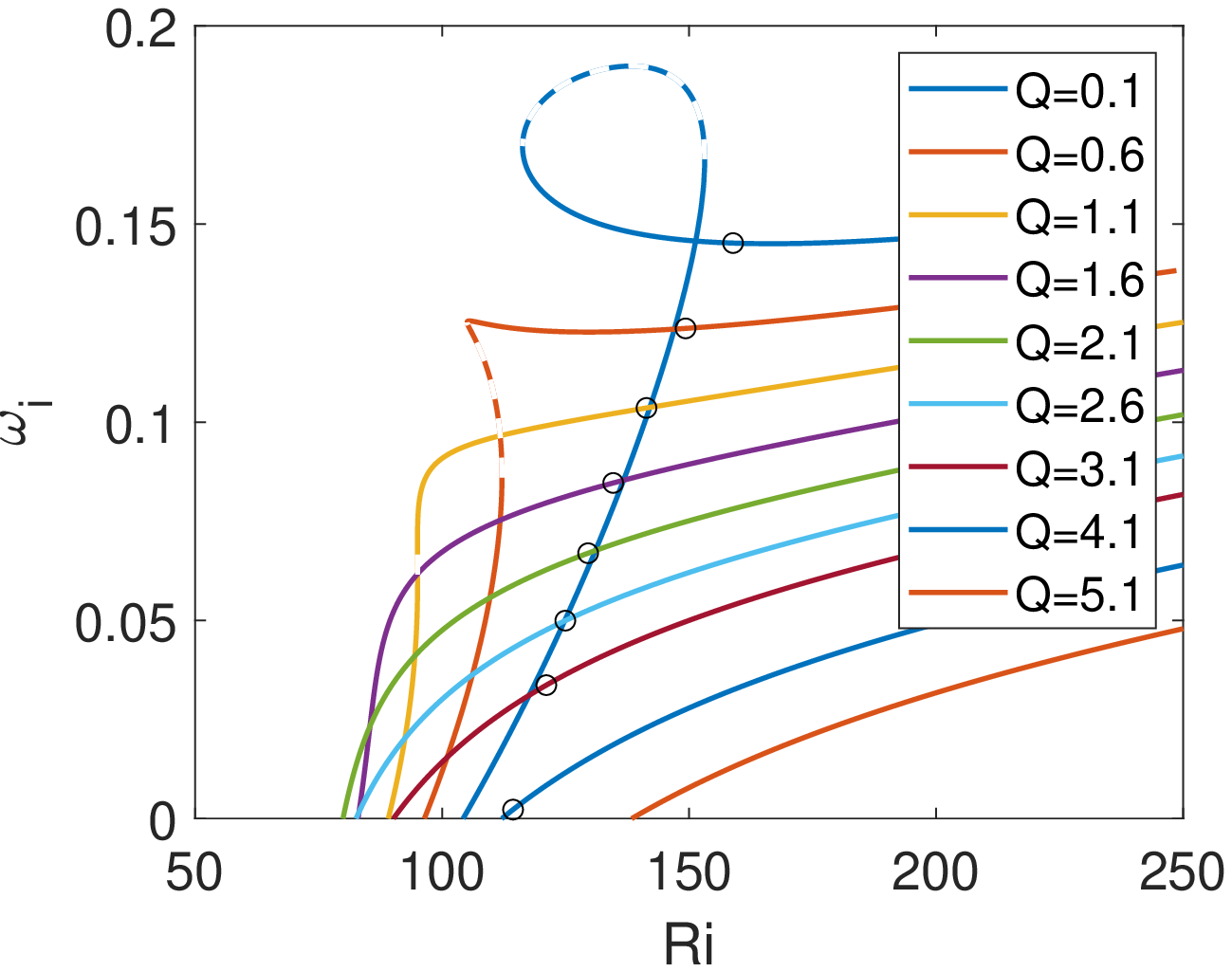}\label{fig:16a}}
	\sidesubfloat[]{\includegraphics[width = 0.45 \columnwidth]{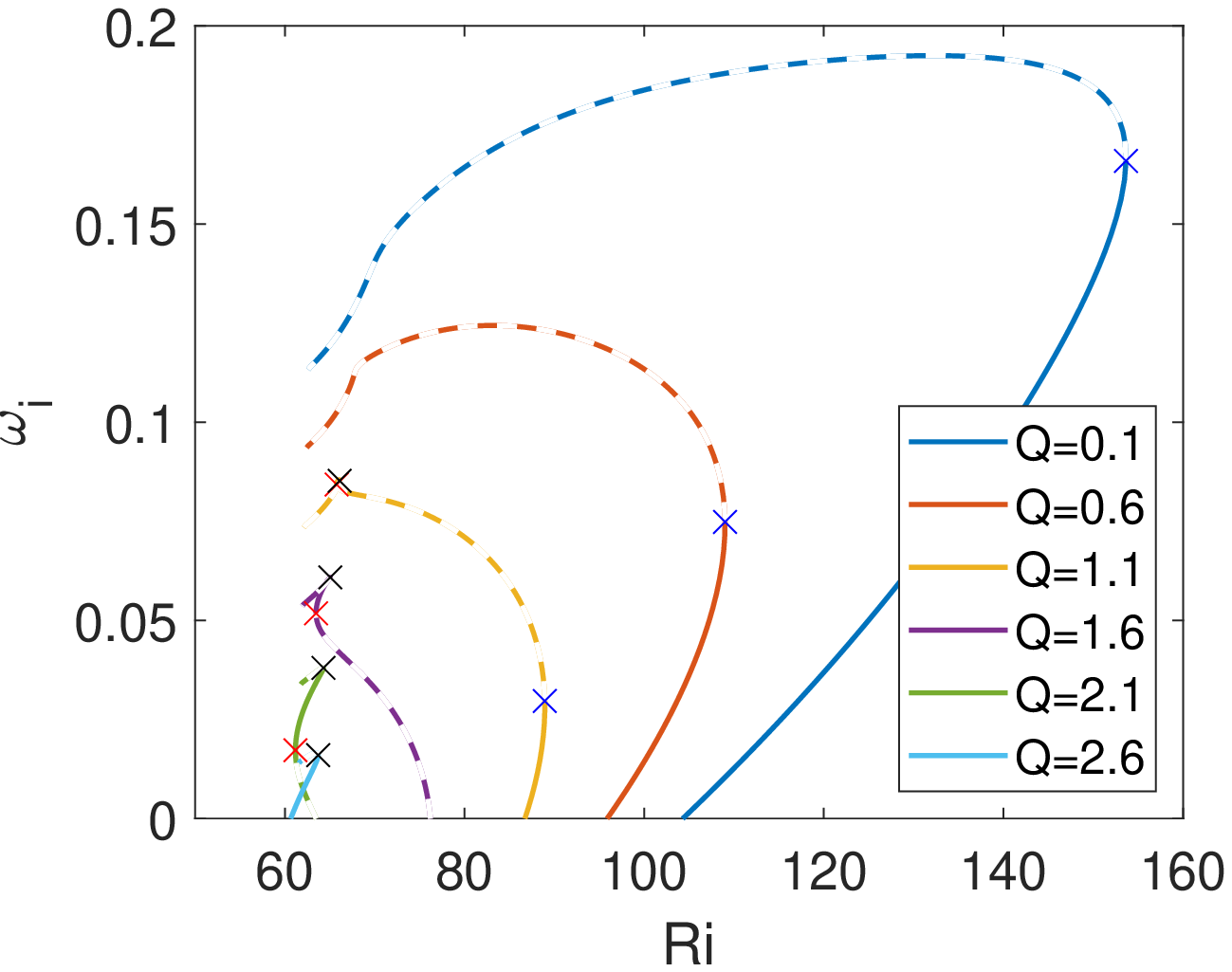}\label{fig:16b}}
\caption{Stability of the non-axisymmetric mode: maximum growth rate $\omega_{i,max}$ at each $Ri$ and $Q$, using $(a)$ model F; and $(b)$ model G. In both figures, \protect \solid, stable; \protect \dashed, unstable to streamwise-uniform perturbation. In $(a)$, the point with maximum $N(0)$ is marked with a black circle in each curve. In $(b)$, the blue, red and black crosses (x) indicate the first, second and third saddle-node points respectively.\label{fig:16}}
\end{figure}
Finally, we have computed the stability of the non-axisymmetric mode (sinuous mode). It was found that, at $m=1$, the non-axisymmetric mode is always the most unstable when $\alpha=0$ in both model F and G (i.e. when the perturbation is streamwise-uniform). Figure \ref{fig:16} shows how the maximum growth rate $\omega_{i,max}$ changes with $Ri$ along the continuation curves shown in figure \ref{fig:3a} and \ref{fig:5a} for model F and G respectively. At low $Q(\sim 0.1)$, the base flow becomes unstable to azimuthally varying (and streamwise-uniform) perturbation at a lower $Ri$ than the streamwise varying axisymmetric perturbation in both models, before the continuation reaches the first saddle-node point. For model F, the growth rate peaks at the middle branch, similar to the axisymmetric mode. However, as the solution continued to the upper branch, the non-axisymmetric mode is destabilising again, even though $N(0)$ is decreasing at the upper branch. For model G, the non-axisymmetric mode stabilises after the first saddle-node point despite increasing $N(0)$.

As $Q$ increases, $\omega_{i,max}$ obtained for all the steady solutions along the continuation decreases. At higher $Q(\gtrsim 1)$, for model F,  $\omega_{i,max}$ remains increasing along the branch continuation even though $N(0)$ started decreasing (as shown by the point indicating maximum $N(0)$ along each continuation curve). For model G, because of the strong stabilising effect of the flow rate, the mode is only unstable when $Q \lesssim 3.0$. In figure \ref{fig:16b}, at higher $Q(\gtrsim 1)$, the mode is destabilising with increasing $N(0)$ before the third saddle-node point. The growth rate $\omega_{i,max}$ peaks at the third saddle-node point. Beyond the third saddle-node point where $N(0)$ tends toward infinity (i.e. approaching the singularity at $Ri_s$), $\omega_{i,max}$ decreases along the continuation curve. This is also the regime where there is a self-similar profile at the centre of the pipe, as discussed in \S\ref{subsec:Bifur_GTD_K}, implying that the non-axisymmetric mode is stabilising when the self-similar plume structure becomes narrower.

The fact that the mode is most unstable when $\alpha=0$ and that it is not as correlated to $N(0)$ as the axisymmetric mode strongly suggest that the non-axisymmetric mode is driven by a different mechanism. In fact, \cite{HP2014b} showed that sinuous mode is driven by gyrotactic instability, even though \cite{HP2014b} did not take into account the spanwise variation. 
According to \cite{HP2014b}, the gyrotactic instability is expected to originate from the following simplified process:
\begin{equation}\label{eq:gyro_mechanism}
    \frac{\partial n'}{\partial t} \sim -  \left ({\esth{r}'} \pardr{N} + N \pardr{{\esth{r}'}}+N \frac{\esth{r}'}{r}+N \pardz{\esth{z}'}+\frac{N}{r}\pardpsi{\esth{\psi}'} \right),
\end{equation}
which also corresponds to the first term and the second line of (\ref{eq:n_pert}). We note that this gyrotactic instability mechanism is different from that of (\ref{eq:instm}), which is driven by the gradient in the base flow average cell swimming. Although both mechanisms originate from gyrotaxis, the former is driven directly by gyrotaxis, while the latter is the result of the net flux of cells in the radial direction due to non-uniform shear rate. A more in-depth discussion on both mechanisms can be found in \cite{HP2014b}. 

To confirm this mechanism, we have performed the stability analysis with the second line of (\ref{eq:n_pert}) suppressed. The non-axisymmetric mode is found no longer unstable for all the parameter space for both model F and G, which shows that the mode is indeed driven by the gyrotactic mechanism. 

\section{Summary and discussion \label{sec:Discussion}}
Thus far, we have explored the bifurcation and stability of a downflowing gyrotactic microorganism suspension in a vertical pipe flow. This work probably provides an almost full analytical picture (bifurcation and stability of steady basic state) for the original experiment of \cite{Kessler1985a,Kessler1985b,Kessler1986,Kessler1986b} with the most up-to-date continuum models (model G, in particular), while extending the stability analysis of \cite{HP2014b} for channel to pipe. In particular, both model F and G have been used in the present study, offering some useful physical insights into the benefits and drawbacks of the existing continuum descriptions. 

\subsection{Model F and G \label{sec:F_G_Comparison}}

The basic-state steady solutions from model F and G have been compared in a series of previous experimental and numerical studies \citep{Bearon2012,Croze2013,Croze2017}. In the present study, a bifurcation analysis has been performed and revealed a complete description of the existence of multiple solutions, their mutual relations and the existence boundaries for the first time. In a stationary suspension, both model F and G exhibit a transcritical bifurcation with $Ri$  (figure \ref{fig:2}). With the addition of a small flow rate $Q$, this transcritical bifurcation with $Ri$ evolves into an imperfect bifurcation involving a saddle-node point (figure \ref{fig:2}). The further increase of the flow rate results in the disappearance of the saddle-node point, exhibiting a cusp bifurcation in terms of two parameters, $Ri$ and $Q$ (figures \ref{fig:3b} and \ref{fig:5b}).

Despite the qualitative similarity in the behaviour of the steady solutions of model F and G, especially at low flow rates, they also exhibit several vital differences. These differences between model F and G stem fundamentally from how the translational diffusivity $\mathsfbi{D}_m$ changes with $S$, especially when $S$ is high (see figure \ref{fig:7b}). The difference in $D_{rr}(S)$ results in different bifurcation behaviour, as shown by the comparison between figure \ref{fig:3a} and \ref{fig:5a}. The differences are particularly profound at high $N(0)$, where the background shear rate $S$ is large. In model F, the plume structure of the steady solution eventually smooths out as it is continued from the middle to the upper branch. This should be related to the recovery of $D_{rr}$ in model F on increasing $S$. In contrast, the monotonically decreasing $D_{rr}(S)$ in model G causes the plume structure of the steady solution to be more focused as it is continued to the upper branch. This is also the essential reason why model G does not admit any steady solution, especially for high $Ri$. On the contrary, in model F, there always exists at least one steady solution, albeit not physically realistic.

As for the stability of the basic state, the main difference between the two models is that model F shows restabilisation of the axisymmetric mode on increasing $N(0)$, whereas model G exhibits a rapid increase in its growth rate (figures \ref{fig:13} and \ref{fig:15}). Given that the only difference between model F and G is in the expression of the translational diffusivity, this difference should also originate from the diffusivity. It has been shown both numerically \citep{Croze2013} and experimentally \citep{Croze2017} that the diffusivity prediction of model F is not as accurate as model G in modelling gyrotactic focusing. This suggests that the restabilisation at high $N(0)$ is likely not physical but an artefact of model F. Furthermore, model F showed that the non-axisymmetric mode has higher growth rate than the axisymmetric mode at high $Ri$, but this is not supported by any of previous experiments, in which only the axisymmetric instability mode (i.e. blip) has been observed. On the contrary, the axisymmetric mode remained the dominant mode in most of the parameter space in model G, which is more consistent with the observation of blips in experiments. These stability results suggest that model F shows little consistency with the experimental observations, and would not be as accurate as model G in predicting the blip occurrence.

\subsection{Limitation and outlook of the continuum descriptions \label{sec:Validity}}
Despite the interesting bifurcation structure and the stability of the basic state, which may offer sound explanations on the previous experiments, care must also be taken in interpreting the present analysis with the assumptions made in modelling of the suspension.
We have assumed in \S\ref{sec:Problem_Formulation} that the contribution of swimming motions of individual microorganisms to the flow field (i.e. the $\bnabla^* \cdot \boldsymbol{\Sigma}_p^*$ term in (\ref{eq:dim_vel})) would be negligible throughout the present study. However, the plume structure of the computed steady solutions exhibits a very high cell density at the pipe axis, implying that this assumption would not be valid near the pipe axis region. Perhaps, a minimal way to address this issue with the continuum models in the present study would be to incorporate the ignored stresslet term. 
To this end, we have recomputed the steady solutions using model G with the stresslet term included, assuming that it is dominated by the contribution of swimming motions of individual microorganisms to the flow field \cite[for a detailed discussion, see also][]{Pedley1990}. The stresslet term considered is given by
\begin{equation}\label{eq:stresslet}
    \boldsymbol{\Sigma}_p^*=n^* T^* (\langle \mathbf{e}\mathbf{e} \rangle - \frac{1}{3} \mathsfbi{I}),
\end{equation}
where $T^*=10^{-10} \mathrm{g cm^2 s^{-1}}$ \citep{Pedley2010b}.

\begin{figure}
\centering{}
	\sidesubfloat[]{\includegraphics[width = 0.45 \columnwidth]{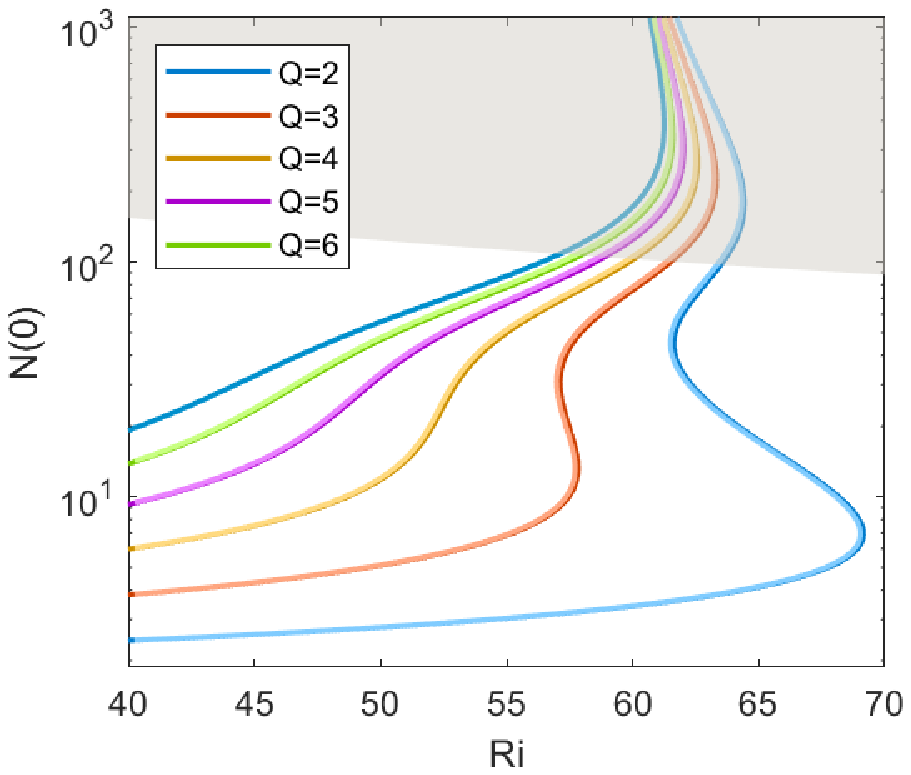}\label{fig:17a}}
	\sidesubfloat[]{\includegraphics[width = 0.45 \columnwidth]{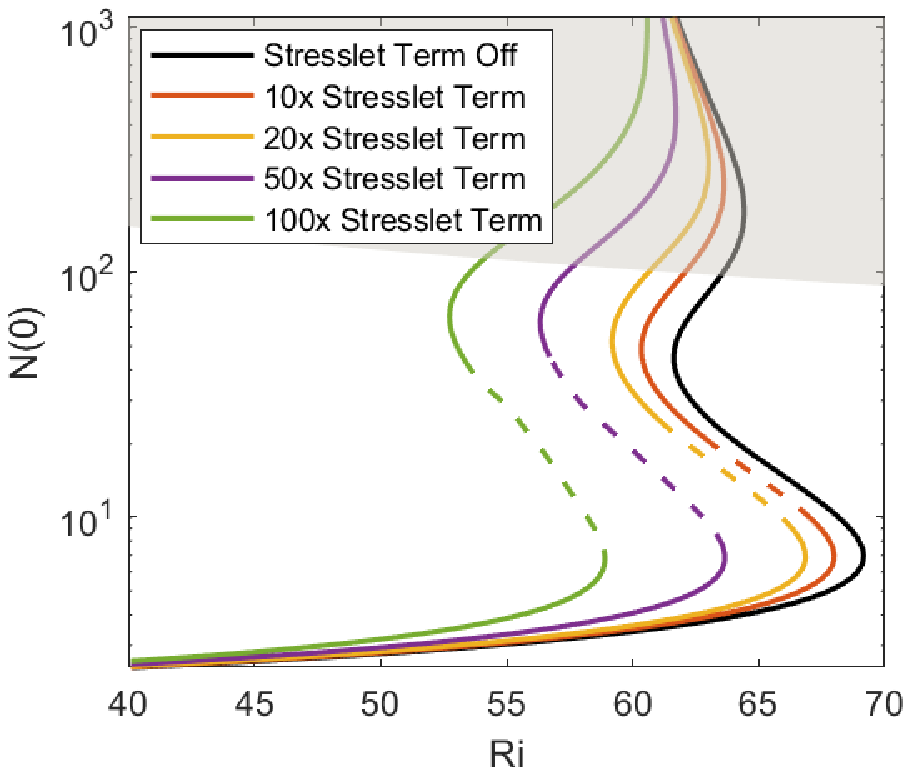}\label{fig:17b}}
\caption{Bifurcation of steady basic-state solution $(a)$ with (lighter lines) and without (darker lines) the stresslet for several $Q$ and $(b)$ for several arbitrarily scaled stresslet strengths $T$ at $Q=2$. The grey area indicates the cases where the volume fraction at the pipe axis is greater than $2.5\%$. In $(a)$, the lighter and darker lines almost overlap. In $(b)$, the dashed line (\protect \dashed) represents the solutions obtained with large residue ($>10^{-4}$) in the solver. }\label{fig:17}
\end{figure}

Figure \ref{fig:17} shows the effect of the stresslet term on the bifurcation of the steady solutions computed with model G. The stresslet term only slightly modifies how $N(0)$ changes with $Ri$, without significant impact on the result (figure \ref{fig:17a}). We have also arbitrarily increased $T$ to further understand any potential role of the stresslet in the bifurcation (figure \ref{fig:17b}). As shown at the top of figure \ref{fig:17b}, the changes in the stresslet strength do not alter the singular behaviour of $N(0)$ (i.e. $N(0) \rightarrow \infty$ ). Moreover, the stresslet term seems to appear to introduce further bifurcation probably involving the emergence of unsteady solutions, as the Newton solver failed  to converge within the prescribed residue ($>10^{-4}$) in the parameter space where the middle-branch solutions reside (dashed lines in figure \ref{fig:17b}). However, this issue has not been further pursued in the present study as the singular behaviour of $N(0)$ is still found to exist even with such a unphysically strong stresslet, implying that the removal of this singularity requires a more sophisticated modelling effort. 

It should be mentioned that, even with the stresslet term included, the  \added{interactions between the cells and the wall and the} near-field interactions between cells are still neglected in the current continuum models. \added{Cell-wall interactions may well be negligible in the present study as the cells tend to swim away from the wall. However, an accurate description of such interactions \citep[e.g.][]{Berke2008,Spagnolie2012,Elgeti2013a,Ezhilan2015a,Bearon2015,Vennamneni2020} may become important when there is an upflow at the centre and cells tend to move towards the wall. In particular, the current no-flux boundary condition in both model F and G does not incorporate the swimming behaviour near the wall, because it simply extends the averaged swimming velocity and the diffusivity obtained without the influence of the wall to the near-wall region. Recently, this issue was comprehensively addressed in by \cite{Jiang2019} with full Smoluchowski equation. As for cell-cell interactions,} \deleted{In other words,} the model with the stresslet term may well be invalid near the centre of the pipe where the cells are highly concentrated. \added{It is known that in the semi-dilute regime, the volume fraction have significant impact on the translational diffusivity \citep[e.g.][]{Hernandez-Ortiz2005,Ishikawa2007,Mehandia2008,Underhill2008}.} To address this issue, we have highlighted the parameter space, where the modelling assumption would break down, in grey in figure \ref{fig:5a} and \ref{fig:17}. The greyed-out areas represent the parameter regime in which the volume fraction at the pipe axis is higher than a threshold value of $2.5 \times 10^{-2}$. This value was previously shown to yield significant changes to the rotational and translational diffusivities by the near-field interaction between cells \citep{Ishikawa2007}.

\added{In the previous study by \cite{Ishikawa2007} for a suspension of squirmers, it was shown that when  the ambient flow is stationary, the rotational diffusivity $D_R$ increases with the volume fraction of the cells while the translational diffusivity decreases.} Therefore, although $N(0) \rightarrow \infty$ was found at $Ri \rightarrow Ri_s$ due to decreasing $D_{rr}$ at high shear, in reality, the near-field interaction at such a high $N(0)$ is likely to increase the effective rotational \replaced{diffusivity}{diffusion} \added{$D_R$} at the pipe axis\deleted{, presumably preventing the unphysical singularity}. \added{However, in general, in the presence of ambient shear, the issue of how $\esth{r}_0$ and $D_{rr,0}$ would change with $D_R$ has not been very well understood.} In this respect, it would be interesting to see how additional modelling incorporating the observations made in the semi-dilute suspensions would modify the plume structure of the solutions in the future.

Finally, besides the violation of the dilute-suspension assumption, the high cell concentration at the centre also gives rise to a rapidly changing shear rate in the plume structure of the solutions. This may also break down the assumptions made in model G. In the application of the GTD theory outlined by \cite{Frankel1991,Frankei1993} and \cite{Manela2003}, the shear rate of the flow was approximated to be locally homogeneous at each spatial point without taking into account the effect of inhomogeneity in the shear. In particular, \cite{Bearon2015} has shown that even in the absence of \added{gyrotaxis that causes the}\deleted{of} net swimming \added{towards the pipe axis}\deleted{(i.e. $\eavg=0$)}, swimmers may still aggregate in the regions of high shear due to the inhomogeneity in the flow shear. In effect, such inhomogeneous shear can also generate an extra net advective cell-concentration flux towards the centre. \added{\cite{Vennamneni2020} categorised such a phenomenon as low-shear trapping}, in which the phenomenological model G fails to capture \added{due to its quasi-homogeneous assumption}. Given that the formation of the upper-branch solutions in the present study is strongly linked to the high shear rate near the centre, this effect might also be significant. 

\subsection{\added{Implications to experimental observations} \label{sec:Implication}}
\added{While the mode structure from the stability analysis is reminiscent of the blips observed in experiments, one needs to be cautious in comparing the experimental observation with the result from the current analysis, performed primarily with the `steady' base-flow solution under the dilute-suspension assumption.
In \S\ref{subsec:Bifur_GTD_K}, we have demonstrated that the solution blows up as $\gamma=\eta Ri \Rey/8 \rightarrow 1$, irrespective of the flow rate $Q$.  We note that $\eta$ is strictly a parameter from the given cell properties and $ReRi$ is proportional to $N^* (h^*)^2$ if the biological properties of the cell and fluid viscosity are fixed. Here, we can define the cross sectional area of the pipe as $A^*= (h^*)^2\pi$.  Therefore, the physical interpretation of the singularity at $\gamma \rightarrow 1$ is that there exist a maximum cell number per unit length of the pipe  $N^* A^*$ which the self-focused steady plume (upper branch) can hold. 

Interestingly, the idea of having a limited capacity in the cell number per unit length of pipe of the steady plume has also been discussed in \cite{Kessler1986}. In the present study, we have extended the theory by demonstrating that if the flow rate is low enough with the centreline velocity at the order of the swimming speed, there exists a lower-branch state which can surpass such a threshold. However, as shown in figures \ref{fig:5a} and \ref{fig:9}, the upper-branch state still has the cell capacity threshold given by $\gamma =1$, implying that there exist a set of certain initial conditions which never reach any of the steady states within the framework of model G. Meanwhile, if the flow rate is high enough, the hysteresis disappear (figures \ref{fig:5a} and \ref{fig:9}) and a steady solution given in this form is no longer possible beyond the cell capacity threshold.
	
In both the experiments of \cite{Kessler1986} and \cite{Denissenko2007} where blips are observed, the total number of cells (per unit length) $N^* A^*$ is an order of magnitude higher than the capacity threshold. Therefore, no steady solution is obtained with model G in the regimes studied in \cite{Kessler1986} and \cite{Denissenko2007}, not allowing for any direct comparison with the experiment.
However, model G and the dilute-suspension assumption remain to be valid for the experiment of \cite{Croze2017}, where the range of parameters did fall within the regime where steady solutions can be found. Unfortunately, in \cite{Croze2017}, the flow profile and the blips are not directly measured.}

Perhaps the most robust feature of the steady solutions found in the present study would be the existence of a cusp bifurcation which emerges in both model F and G. The cusp bifurcation involves bistability of the steady solutions. In such a system, the asymptotic state is highly sensitive to the form of the given initial condition and is often featured with a hysteresis which involves a discontinuous change of the state upon a continuous change of the bifurcation parameter ($Ri$ in this case).
Such sensitivity might explain why the experimental observations made for the blip formation in the $Q-Ri$ space are so scattered and qualitatively different from the prediction of \cite{HP2014b}  \citep[see figure 5 in][]{Croze2017}.
\deleted{The `discontinuous change' of the plume structure from the lower to the upper branch or vice versa could also be potentially related to the emergence of a `train-like' thread observed in Denissenko \& Lukaschuk (2007). In their experiment, the `train-like' thread originated at the top of the pipe and propagated downstream much faster than normal blips, reminiscent of a sudden switch of the state to the upper branch (which has a much higher $U(0)$) from the lower one. While the upper branch may be a potential candidate to explain the `train-like' thread, at least a two-dimensional description involving in the radial and axial directions would be needed to accurately model its spatial-temporal dynamics instead of relying on the over-simplified one-dimensional description in the present study.}

\section*{Acknowledgement}
L. F. gratefully acknowledges funding from the President's PhD Scholarship of Imperial College London. We would like to thank Dr O. Croze and Dr D. Jin for sharing their experimental observations and providing their numerical solver to verify our results. 

\section*{Declaration of interests}
The authors report no conflict of interest.

\appendix
\section{Linearised equations for perturbed average swimming orientation vector and diffusivity \label{sec:lin_diff}}
In (\ref{eq:n_pert}), the values of $\diffusion'_m$ and $\eavg'$ are required. To obtain these values, we first need to compute $f(\mathbf{e})'$ with a perturbation of $\boldsymbol{\Omega}'$:
\begin{equation} \label{eq:pdf_p}
    \bnabla_e \bcdot \left[ \lambda[\mathbf{e_2}-(\mathbf{e_2} \bcdot \mathbf{e})\mathbf{e}]f'+\frac{1}{2 D_R} \boldsymbol{\Omega} \wedge \mathbf{e} f' \right]  -\bnabla^2_e f' = -\bnabla_e \bcdot \left[ \frac{1}{2 D_R} \boldsymbol{\Omega}' \wedge \mathbf{e} f \right].    
\end{equation}{}
Then, $\eavg'$ can be computed with $f(\mathbf{e})'$ by
\begin{equation}
\langle \mathbf{e} \rangle' =\int_{\|\mathbf{e}\|=1} ~\mathbf{e} f'(\mathbf{e})~d^2\mathbf{e}.
\end{equation}
For model F, $\diffusion'_F$ is obtained easily by \citep[][]{HP2014b}
\begin{equation}
   \diffusion_F'=\tau (\langle \mathbf{e} \mathbf{e}\rangle' -\langle \mathbf{e} \rangle \langle \mathbf{e} \rangle'-\langle \mathbf{e} \rangle \langle \mathbf{e} \rangle').
\end{equation}
However, for model G, the process is more involved, as $\diffusion'_G$ not only depends on $\boldsymbol{\Omega}'$, but also $\mathsfbi{G}'=\bnabla \mathbf{u}'$. Hence,
\begin{equation}
   \diffusion_G'= \int_{\|\mathbf{e}\|=1}  \left[ \mathbf{b}'\mathbf{e}
   +\frac{\mathbf{b}'\mathbf{b}}{f} \cdot \mathsfbi{G}
   +\frac{\mathbf{b}\mathbf{b}'}{f} \cdot \mathsfbi{G}
   -\frac{\mathbf{b}\mathbf{b}f'}{f^2} \cdot \mathsfbi{G}
   +\frac{\mathbf{b}\mathbf{b}}{f} \cdot \mathsfbi{G}'
   \right]^{sym} d^2 \mathbf{e},
\end{equation}
where the perturbed $\mathbf{b}'$ due to $\mathsfbi{G}'$ (and $\boldsymbol{\Omega}'$) is needed. $\mathbf{b}'$ can be computed by solving
\begin{equation} \label{eq:GTD_bp}
    \bnabla_e \cdot \left[ \mathbf{\dot{e}}\mathbf{b}'- \bnabla_e\mathbf{b}'\right]-\mathbf{b}' \cdot \mathsfbi{G}= -\eavg'f+(\mathbf{e}-\eavg)f'+\mathbf{b} \cdot \mathsfbi{G}'-\bnabla_e \bcdot \left[ \frac{1}{2 D_R} (\boldsymbol{\Omega}' \wedge \mathbf{e}) \mathbf{b} \right].
\end{equation}
In practice, the left-hand side of (\ref{eq:pdf_p}) and (\ref{eq:GTD_bp}) is the same linear operator used in  (\ref{eq:pdf}) and (\ref{eq:GTD_b}), while their right-hand side can be viewed as different forcing terms. Therefore, $f'$ and $\mathbf{b}'$ can be obtained by imposing the different forcing term on the right-hand side of (\ref{eq:pdf_p}) and (\ref{eq:GTD_bp}), similarly to the framework of \cite{HP2014a,HP2014b}. 

\section{Equations for linear stability}\label{sec:lin_modal}
Using the framework in Appendix \ref{sec:lin_diff}, $\eavg'$ and $\diffusion_m'$ can be written as linear combinations of the components of $\boldsymbol{\Omega}'$ (Model F and G) and $\mathsfbi{G}'$ (Model G), hence also as a linear combination of $\mathbf{u}'$. This allows us to write $\eavg'$ and $\diffusion_m'$ as follows:
\begin{subequations}
\begin{eqnarray}
\eo{r}' & = & \frac{\xi_1}{D_R}(\pardz{v'}-\pardr{u'}); \\
\eo{z}' & = &  \frac{\xi_2}{D_R}(\pardz{v'}-\pardr{u'}); \\
\eo{z}' & = &  \frac{\xi_3}{D_R}(\frac{1}{r}\pardpsi{u'}-\pardz{w'})+\frac{\xi_4}{r D_R}(\pardr{r w'}-\pardpsi{v'}); 
\end{eqnarray}
\begin{eqnarray}
D_{rr}' & = &\frac{1}{D_R}(\xi_5\pardr{u'}+\xi_6\pardr{v'}+\xi_7\pardr{w'}+\xi_8\pardz{u'}+\xi_9\pardz{v'}+\xi_{10}\pardz{w'} \nonumber \\ 
& + & \frac{1}{r}(\xi_{11}\pardpsi{u'}+\xi_{12}(\pardpsi{v'}-w')+\xi_{13}(\pardpsi{w'}+v'))); \\
D_{rz}' & = &\frac{1}{D_R}(\xi_{14}\pardr{u'}+\xi_{15}\pardr{v'}+\xi_{16}\pardr{w'}+\xi_{17}\pardz{u'}+\xi_{18}\pardz{v'}+\xi_{19}\pardz{w'} \nonumber \\ 
& + & \frac{1}{r}(\xi_{20}\pardpsi{u'}+\xi_{21}(\pardpsi{v'}-w')+\xi_{22}(\pardpsi{w'}+v'))); \\
D_{r\psi}' & = &\frac{1}{D_R}(\xi_{23}\pardr{u'}+\xi_{24}\pardr{v'}+\xi_{25}\pardr{w'}+\xi_{26}\pardz{u'}+\xi_{27}\pardz{v'}+\xi_{28}\pardz{w'} \nonumber \\ 
& + & \frac{1}{r}(\xi_{29}\pardpsi{u'}+\xi_{30}(\pardpsi{v'}-w')+\xi_{31}(\pardpsi{w'}+v'))), 
\end{eqnarray}
\end{subequations}
where $\xi_{1-4}$ are the same for model F and G, but $\xi_{5-31}$ are different for model F and G.

Application of the normal-mode assumption of (\ref{eq:normal_mode}) to (\ref{eq:lin_full}), we get
\begin{subequations}\label{eq:lin_full_modal}
    \begin{equation}
        i \alpha{\hat{u}}+\frac{1}{r}\pardr{r\hat{v}}+\frac{1}{r}i m{\hat{w}}=0,
    \end{equation}
    \begin{equation}
        i \omega \hat{u}+ L_{OS}{\hat{u}}+ \pardr{U} \hat{v}=-i \alpha{\hat{p}} + \Ri~ \hat{n},
    \end{equation}
    \begin{equation}
        i \omega \hat{v} + L_{OS}{\hat{v}}  =-\pardr{\hat{p}}+\frac{1}{\Rey}( -\frac{\hat{v}}{r^2} - \frac{2 i m}{r^2}{\hat{w}}),
    \end{equation}
    \begin{equation}
        i \omega \hat{w} + L_{OS}{\hat{w}} =-\frac{1}{r}i m{\hat{p}}+
        \frac{1}{\Rey}(- \frac{\hat{w}}{r^2} +\frac{2 i m}{r^2}{\hat{v}}),
    \end{equation}
    \begin{eqnarray}
        i \omega{\hat{n}} &+& L_n \hat{n} \nonumber \\
        & + & 
            \pardr{N}\hat{v} + (\pardr{N} + N \pardr{}+\frac{N}{r})(\frac{i \alpha \xi_1}{D_R}{\hat{v}}-\frac{\xi_1}{D_R} \pardr{\hat{u}}) \nonumber \\
        & + & i \alpha N \frac{\xi_2}{D_R}(i \alpha{\hat{v}}-\pardr{\hat{u}})+\frac{i m N}{r}(\frac{\xi_3}{D_R}(\frac{i m}{r}\hat{u}-i \alpha \hat{w})+\frac{\xi_4}{r D_R}(\pardr{r \hat{w}}-i m \hat{v})) \nonumber \\
        & = &
            \frac{1}{D_R^2} \left[ (\frac{1}{r}\pardr{N}+\pardr{N}\pardr{}+\pardrr{N}) \left( \xi_5\pardr{\hat{u}}+\xi_6\pardr{\hat{v}}+\xi_7\pardr{\hat{w}} \right. \right. \nonumber \\ 
        & + &
            \left. i \alpha (\xi_8{\hat{u}}+\xi_9{\hat{v}}+\xi_{10}{\hat{w}})+\frac{i m}{r}(\xi_{11}{\hat{u}}+\xi_{12}{\hat{v}}+\xi_{13}{\hat{w}})+\frac{1}{r}(-\xi_{12}\hat{w}+\xi_{13}\hat{v}) \right) \nonumber \\ 
        & + & 
            {i \alpha} \pardr{N} \left( \xi_{14}\pardr{\hat{u}}+\xi_{15}\pardr{\hat{v}}+\xi_{16}\pardr{\hat{w}}+i \alpha (\xi_{17}\hat{u}+\xi_{18}\hat{v}+\xi_{19}\hat{w}) \right. \nonumber \\ 
        & + & 
            \left. \frac{i m}{r}(\xi_{20}\hat{u}+\xi_{21}\hat{v}+\xi_{22}\hat{w})+\frac{1}{r}(-\xi_{21}\hat{w}+\xi_{22}\hat{v}) \right) \nonumber \\ 
        & + & 
            \frac{i m}{r} \pardr{N} \left( \xi_{23}\pardr{\hat{u}}+\xi_{24}\pardr{\hat{v}}+\xi_{25}\pardr{\hat{w}}+i \alpha (\xi_{26}\hat{u}+\xi_{27}\hat{v}+\xi_{28}\hat{w}) \right. \nonumber \\ 
        & + & 
            \left. \left. \frac{i m}{r}(\xi_{29}\hat{u}+\xi_{30}\hat{v}+\xi_{31}\hat{w})+\frac{1}{r}(-\xi_{30}\hat{w}+\xi_{31}\hat{v}) \right) \right],
    \end{eqnarray}
    where 
    \begin{equation}
        L_{OS}= i \alpha U -\frac{1}{Re}\left( \frac{1}{r}\pardr{} ( r \pardr{}) -\alpha^2-\frac{m^2}{r^2} \right),
    \end{equation}
    and
    \begin{eqnarray}
        L_n & = & (\frac{\eo{r}}{r}+\pardr{\eo{r}})+\eo{r}\pardr{} +i \alpha U + i \alpha \eo{z} +\frac{i m \eo{z}}{r} \nonumber \\
        & - &
        \frac{1}{D_R} \left[ \frac{1}{r} \left(D_{rr,0}\pardr{}+ i \alpha D_{rz,0}+2 i m D_{r\psi,0}\pardr{}+i m \pardr{D_{r\psi,0}}-2 \alpha m D_{\psi z,0} \right)  \right. \nonumber \\
        & + &
            \pardr{D_{rr,0}}\pardr{}+2 i \alpha \pardr{D_{rz,0}}+i \alpha D_{rz,0}\pardr{} \nonumber \\
        & + &
            \left. D_{rr,0}\pardrr{}-\alpha^2 D_{zz,0} -\frac{m^2}{r^2}D_{\psi \psi,0}
             \right].
    \end{eqnarray}
    The boundary condition at the wall is
    \begin{equation}\label{eq:velp_BC_modal}
        \hat{u}|_{r=1} =\hat{v}|_{r=1}=\hat{w}|_{r=1}= 0
    \end{equation}
    and
    \begin{equation}\label{eq:np_BC_modal}
      N \hat{v}+ N \esth{r}' + \eo{r} \hat{n}=\frac{1}{D_R}
      \left(D_{rr}' \pardr{N}+D_{rr,0} \pardr{\hat{n}}+i \alpha D_{rz,0} \hat{n}+i m \frac{D_{r\psi,0}}{r}\hat{n} \right).
    \end{equation}
    The compatibility conditions at the centre of the pipe are
    \begin{eqnarray}
        \hat{u}=\hat{v}=\hat{w}=\hat{p}=\pardr{\hat{n}}=0 & \mathrm{when} & m \geq 2; \\
        \hat{v}+i\hat{w}=0, \hat{u}=\pardr{\hat{n}}=\hat{p}=0 & \mathrm{when} & m = 1; \\
        \hat{v}=\hat{w}=0, \pardr{\hat{u}}=\pardr{\hat{n}}=0 & \mathrm{when} & m = 0.
    \end{eqnarray}
\end{subequations}
These equations can now be discretised in the radial direction and solved as an eigenvalue problem, as mentioned in \S \ref{sec:numerical}.

\bibliography{references}
\bibliographystyle{jfm}

\end{document}